\documentclass[twocolumn,pre,aps,nofootinbib]{revtex4}
\usepackage[latin9]{inputenc}
\usepackage{float}
\usepackage{amsmath}
\usepackage{amssymb}
\usepackage{graphicx}
\usepackage{wasysym}

\makeatletter

\newcommand{\lyxdeleted}[3]{}

\@ifundefined{textcolor}{}
{%
 \definecolor{BLACK}{gray}{0}
 \definecolor{WHITE}{gray}{1}
 \definecolor{RED}{rgb}{1,0,0}
 \definecolor{GREEN}{rgb}{0,1,0}
 \definecolor{BLUE}{rgb}{0,0,1}
 \definecolor{CYAN}{cmyk}{1,0,0,0}
 \definecolor{MAGENTA}{cmyk}{0,1,0,0}
 \definecolor{YELLOW}{cmyk}{0,0,1,0}
}

\usepackage{amsfonts}
\providecommand{\U}[1]{\protect \rule{.1in}{.1in}}

\makeatother

\begin{document}
\author{Miodrag L. Kuli\'{c}}
\title{$\mathrm{\textrm{\textrm{\ensuremath{\textrm{H}_{3}}S}}}$ is a high-$\kappa$
superconductor with columnar pinning defects}
\affiliation{Institute for Theoretical Physics, Goethe-University Frankfurt am
Main, Germany }
\begin{abstract}
Recently, the existence of superconductivity in $\mathrm{H_{3}S}$
and other high-pressure hydrides is called into question, because
of some flawed magnetic measurements. In Ref. \cite{Kulic-RTSC-v1})
the SCPC model is proposed, where strong pinning of vortices is due
to long columnar defects in $\mathrm{H_{3}S}$ - with lengths $\mathrm{L}$
of the order of vortex lengths $\mathrm{L_{v}}$. Two relevant type
of experiments in magnetic fields are explained by this model: (1)
Reduction (with respect to standard superconductors) of the thermal
broadening of resistance ($\mathit{TBR}$) in magnetic field $\mathrm{h=H/H_{c2}}$,
$\mathrm{\mathrm{\delta t_{c}(h)}}$, is governed by the small parameter
$\mathrm{C=}$$\mathrm{\xi_{0}}/\mathrm{L}$, i. e. $\mathrm{\mathrm{\delta t_{c}^{scpc}(h)}\sim C^{1/2}h^{1/2}}$
with $\mathrm{C\sim10^{-3}}$. This gives $\mathrm{\mathrm{\delta t_{c}^{scpc}(h)}\apprle0.01}$
for $\mathrm{h\lesssim0.01}$ and $\mathrm{L}\sim1$ $\mu m$, which
is in satisfactory agreement with measurements in $\mathrm{H_{3}S}$
\cite{Eremets-Meissner-2022}. The $\mathit{TBR}$ measurements give,
that in $\mathrm{H_{3}S}$ there is a large shift of the irreversible
line $\mathrm{B_{irr}(T)}$ towards the $\mathrm{B_{c2}(T)}$ line,
with $\mathit{\mathrm{B_{ir}^{(H_{3}S)}\sim C^{-1}(1-t)^{2}}}$ instead
of $\mathrm{B_{ir}\sim(1-t)^{3/2}}$ ($\mathrm{t=T/T_{c}}$) - in
standard superconductors with point defects. (2) In ZFC (zero field
cooled) experiments on penetration of the magnetic field ($\mathit{PMF}$)
in $\mathrm{H_{3}S}$ \cite{Troyan}, the latter reaches the center
of a superconducting disk at much larger external fields, i. e. for
$\mathrm{B_{0}\gg B_{c1}}$. The later is due to the pronounced pinning
of vortices, but not to the Meissner effect. The calculated $\mathrm{T}$-
dependence of the penetrated, perpendicular $\mathrm{B_{\perp}(T)}$
and parallel $\mathrm{B_{\parallel}}(\mathrm{T})$, magnetic field
into the sample is in satisfactory agreement with the experimental
results for $\mathrm{H_{3}S}$ in \cite{Troyan}, where $\mathrm{B_{\perp}>B_{\parallel}}$.
Problems related to measurements of the Meissner effect in $\mathrm{H_{3}S}$
are also discussed. The SCPC model, applied on the measurements in
the bulk $\mathrm{H_{3}S}$ sample, predicts, that the latter is a
high-$\kappa$ superconductor with $\xi_{0}\approx(15-20)$ $\textrm{Å}$
, $\lambda_{0}\approx(1-2)\times10^{3}$$\textrm{Å}$ ; $\kappa\approx(50-100)$,
$\mathrm{\mathrm{\mu_{0}}H_{c1}(0)\approx}(18-60)$ $\mathrm{m\mathrm{T}}$,
$\mathrm{\mathrm{\mu_{0}}H_{c0}}$$\approx(0,6-1,1)$ $\mathrm{\mathrm{T},}$
$\mathrm{\mathrm{\mu_{0}}H}_{c2}(0)\approx(80-140$) $\mathrm{T}$. 
\end{abstract}
\date{\today}
\maketitle

\section{Introduction }

The first almost room-temperature superconductor was reported in 2015
in sulphur-hydrides ($H_{3}S$) with $\mathrm{T_{c}}$$\approx203$
$\mathrm{K}$ under high pressure $P\approx150$ $GPa(\approx1.5$
$\mathrm{Mbar})$, which is based on resistivity measurements \cite{Drozdov-2015}.
This finding opens a new frontier in physics and a number of other
$\mathrm{HP}$-hydrides were predicted before these were synthesized
thereafter. Let us mention some of them with $\mathrm{T_{c}}$$\geq200$
$\mathrm{K}$ - such as $\mathrm{LaH_{10}}$with $T_{c}\approx250$
$\mathrm{K}$, $\mathit{P\mathrm{\approx190}}$ $\mathrm{GPa}$ \cite{LaH10};
$\mathrm{LaYH}_{x}$ with $\mathrm{T_{c}}$$\mathrm{=253}$ $\mathrm{K}$,
$\mathrm{P}$$\approx190$ $\mathit{GPa}$ \cite{Semenok}. In all
of them $\mathrm{T_{c}}$ goes down by increasing magnetic field,
compatible with the standard theory of superconductivity. There are
also reports on the room temperature superconductivity in the CSH
hydride - a superconductor based on C, S and H, with $\mathrm{T_{c}}$$\mathrm{=287}$
$\mathrm{K}$ at $\mathrm{P}$$\approx267$ $\mathrm{GPa}$ \cite{Snider}.
However, this result was not yet confirmed, neither experimentally
nor theoretically by other groups. 

There is also theoretical support for the (almost) room-temperature
superconductivity in HP-hydrides, which are based on the calculated
critical temperature $\mathrm{T_{c}}$ in the microscopic Migdal-Eliashberg
theory for superconductivity - which is due to the electron-phonon
interaction. Moreover, the theoretical prediction of superconductivity
in HP-hydrides \cite{Duan} is a rare example in the physics of superconductivity,
that the theory goes ahead of experiments, by predicting $\mathrm{T_{c}(\approx200}$
$\mathrm{K)}$ in $\mathrm{H_{3}S}$. Note, that in \cite{Drozdov-2015}
the $\mathrm{H_{2}S}$ structure is proposed, while in \cite{Mazin-H3S}
it is argued, that at high pressures the phase diagram favors decomposition
of $\mathrm{H_{2}S}$ into $\mathrm{H_{3}S}$ and pure $\mathrm{S}$. 

A main proof for superconductivity is the existence of the Meissner
effect, where the magnetic field $\mathrm{H<H_{c1}}$ - applied above
$\mathrm{T_{c}}$, is $\mathit{expelled}$ from the sample at temperatures
below $\mathit{\mathrm{T_{c}}}$ - the field cooled ($\mathrm{FC}$)
experiment. However, until now the Meissner effect is not proved experimentally
in HP-hydrides, which caused justified criticism on this subject in
\cite{Hirsch-1-1}-\cite{Hirsch-Marsiglio-arXiv V5-2022-1}. The failure
to measure the Meissner effect in $\mathrm{H_{3}S}$ is due to the
following reasons: ($\mathit{i}$) Magnetic measurements in small
samples are delicate; ($\mathit{ii}$) The existence of some extrinsic
paramagnetic effects in samples. In that respect, some contradictory
experimental results and their inconsistent theoretical interpretations
- given in \cite{Eremets-Meissner-2022},\cite{Drozdov-2015},\cite{Minkov-Meissner-2021},
were among the reasons that Hirsch and Marsiglio even called into
question the existence of superconductivity in $\mathrm{HP}$-hydrides
\cite{Hirsch-1-1}-\cite{Hirsch-Marsiglio-arXiv V5-2022-1}. In the
following, we are going to show that some magnetic measurements in
HS can refute this skepticism. 

The content of the article is following. In Section $\mathrm{II}$
the SCPC model - firstly introduced in \cite{Kulic-RTSC-v1} for hard
type-II superconductors with strong columnar pinning centers, is further
elaborated and applied to the magnetic measurements in the $\mathrm{H_{3}S}$
superconductor. This model proposes, that the vortex pinning is due
to long columnar defects ($\textrm{\ensuremath{\mathrm{L\sim L_{v}\gg\xi_{0}}}}$)
- with the radius of the cross-section $\mathrm{r\sim\xi_{0}}$. In
that case, the core and electromagnetic pinning contribute almost
equally to the elementary pinning energy $\mathrm{U_{p}}.$ This is
an optimal situation for pinning in a superconductor, since there
is a maximal gain in the condensation and electromagnetic vortex energy
. This gives a maximal pinning force if the superconductivity is fully
suppressed in the columnar defects, i.e. for $\Delta=0$ inside a
defect. It seems that $\Delta$ is finite in $\mathrm{H_{3}S}$, but
$(\Delta<\Delta_{0})$ and the critical current density is smaller
than the maximal one, i.e. $\mathrm{j_{co}=\mathrm{\eta_{col}j_{c0}^{max}}}$
with $\mathrm{\eta_{col}<1}$. In Section $\mathrm{III}$ the reduction
of the temperature broadening of resistance ($\mathit{TBR}$) in a
magnetic field, $\delta t_{c}^{exp}(h)$ (with $\mathrm{h=H/H_{c2}}$),
in HP-hydrides is studied in the SCPC model and applied to $\mathrm{H_{3}S}$.
The temperature dependence of the irreversibility line with $\mathit{\mathrm{B}_{\mathrm{ir}}\sim\mathrm{C^{-1}(1-t)^{2}}}$
is predicted within the SCPC model also in Section $\mathrm{III}$.
In Section $\mathrm{IV}$ the SCPC model is applied in studying the
penetration of the magnetic field ($\mathrm{\mathit{PMF}}$) into
the center of the $\mathit{\mathrm{H_{3}S}}$ sample. The $\mathrm{\mathrm{PMF}}$
measurements are analyzed in the Bean critical state model, first
for a long superconducting cylinder (the parallel configuration $\mathrm{B_{\parallel}}$approximately
realized in \cite{Troyan} ) and then for thin disks (perpendicular
configuration $\mathrm{B_{\perp}}$). It is shown, that the critical
magnetic field $\mathbf{B_{\mathrm{p}}}$ at which it penetrates into
the center of the sample is larger for the perpendicular geometry
than for the parallel one, i.e. $\mathrm{B_{p\perp}>B_{p\parallel}}$.
The comparison of the theory and the experimental results for $\mathrm{\mathit{PMF}}$
in $\mathrm{H_{3}S}$ \cite{Troyan} gives a large critical current
$\mathrm{j}$$_{c0}$$\apprge10^{7}$ $\mathrm{A/cm^{2}}$.

In Section $\mathrm{\mathrm{I}V}$ we discuss some experimental controversies
related to the Meissner effect and FC experiments in $\mathrm{H_{3}S}$,
which display a large residual paramagnetic magnetization. The latter
fact does not fit into the classical theory of the Meissner effect.
Note, that the paramagnetic signal can be present in some magnetic
superconductors, superconductors with intrinsic magnetic moments in
\cite{Tallon-Ru}-\cite{EuFeIrAs} or extrinsic ones \cite{Pan-FS}.
Finally, in Section $\mathrm{V}$ the obtained results for $\mathrm{TBR}$
and $\mathrm{PMF}$ in the SCPC model for $\mathrm{H_{3}S}$ are summarized
and discussed. Here, the crucial difference between pinning forces
in HTSC-cuprates and HP-hydridesis also discussed. Note, in the following
we use the notation $\xi_{0}\approx\xi(T\ll T_{c})$ and $\lambda_{0}\approx\lambda(T\ll T_{c})$

\section{Strong vortex pinning by columnar defects in H$_{3}$S}

\subsection{Single vortex pinning }

To increase the critical current density it is desirable to have extended
(long columnar) pinning defects - the $\mathit{correlated}$ $\mathit{disorder}$.
In this case the pinning potential is correlated over the extended
size of the defect $\mathrm{L}$$\sim\mathrm{L_{v}}$. ($\mathrm{L_{v}}$
is the vortex length.) This means that the superconductivity is suppressed
in a large volume $\mathrm{V_{col}}$$=\pi\xi_{0}^{2}$$\mathrm{L_{v}}$
and therefore a single vortex prefers sitting on this defect. Note,
that the maximal pinning energy $\mathrm{U_{p}^{max}=\epsilon_{p}^{max}V_{col}}$
is reached when the superconductivity is fully suppressed in the core
of defects, i.e. when $\Delta=0$ in dielectric columnar defects.
This might be not the case in $\mathrm{H_{3}S}$, where the real pinning
energy density is $\mathrm{\epsilon_{p}=\eta_{col}\cdot\epsilon_{p}^{max}}$
with the reducing factor $\mathrm{\eta_{col}}<1$, because the gap
inside the columnar defect is finite, $\Delta<\Delta_{0}$. In that
case $\mathrm{\eta_{col}=\epsilon_{p}/\epsilon_{p}^{max}\sim(1}-\Delta^{2}/\Delta_{0}^{2})$.
As a result, the critical current density for a single vortex (sitting
on such a defect) is given by $\mathrm{j_{c}}=\eta_{\mathrm{col}}\cdot\mathrm{j}_{c}^{\mathrm{max}}$.

Note, that in HTSC-cuprates long columnar pinning defects are made
artificially by irradiating $\mathrm{YBa_{2}Cu_{3}O_{7}}$ single
crystalline samples of small platelets of $\mathrm{1}$$\times1\times0.02$
$\mathrm{mm^{3}}$ with different doses of $\mathrm{580}$ $\mathrm{MeV}$
$\mathrm{^{116}Sn^{30+}}$ ions \cite{Civale}-\cite{Blatter-RMP}.
The density of the ion doses are $\mathrm{D\approx5\times10^{10},}1.5\times10^{11},2.4\times10^{11}$$\mathrm{ions/cm^{2}}$,
which are equivalent to the corresponding vortex densities with the
magnetic induction $\mathrm{B_{\phi}\approx1,3,5}$ $\mathrm{T}$
- $\mathrm{B_{\phi}}$ is the matching field. Since the ionization
energy-loss rate was $\mathrm{2.7}$ $\mathrm{keV/\textrm{Å}}$ they
produce long tracks with the length $\mathrm{L\sim20-30}$$\textrm{ }$$\mu m$
and diameter $\mathrm{\sim50\textrm{ Å}}$. In that case $\mathrm{j_{c}(\sim j_{c0})}$
is of the order $\mathrm{1.5}$$\times10^{7}$$\mathrm{A/cm^{2}}$
at $\mathrm{T=5}$ $\mathrm{K}$ and $\mathrm{10^{6}A/cm^{2}}$at
$\mathrm{T=77}$ $\mathrm{K.}$ These current densities are much larger
than those for point defects, where the pinning is due to oxygen vacancies.
In the columnar case the irreversible line $\mathrm{B_{ir}(T)}$ (the
line below which the pinning is pronounced) lies higher than for weak
pinning with point defects \cite{Civale}. 

In the following it is argued, that the columnar pinning defects dominate
in $H_{3}S$ - see blue colored cylinders in $\mathit{Fig.}$ $\mathit{2}$,
with the averaged distance $\mathrm{d_{\phi}=(\varPhi_{0}/B_{\phi})^{1/2}}$
\cite{Mrktchyan-Scmidt}-\cite{Kulic-CD}. For an optimal pinning
the superconductivity should be fully destroyed inside the dielectric
columnar defect with radius $\mathrm{r\gtrsim\xi}$ \cite{Mrktchyan-Scmidt}-\cite{Kulic-CD}.
In this case, both mechanisms of pinning - the $\mathit{core}$ and
the $\mathit{electromagnetic}$ one, are operative with almost same
pinning energy \cite{Kulic-CD}. The maximal pinning energy is given
by $\mathit{\mathrm{U_{p}}^{\mathrm{scpc}}}$$\approx2\pi\xi^{2}L_{\mathrm{v}}(H_{c}^{2}/8\pi)\equiv L_{\mathrm{v}}\mathrm{\epsilon_{p}^{max}}$
with $\mathrm{\epsilon_{p}^{max}}\approx\Phi_{0}^{2}/32\pi^{2}\mathrm{\lambda^{2}(T)}$.
Since the elementary pinning force (per unit vortex length) $f_{p}=$$\mathit{\epsilon_{p}^{\mathrm{max}}/\xi}$
is balanced in the critical state by the Lorenz force (per unit length)
$\mathit{\mathrm{f_{L}(\equiv F_{L}/L_{v})=j_{c}\Phi_{0}/c}}$, the
critical current density $\mathrm{j_{c}^{max}}$ in the $\mathrm{SCPC}$
model is given by \cite{Mrktchyan-Scmidt}-\cite{Kulic-CD}
\begin{equation}
j_{c}^{\mathrm{max}}\approx\frac{c\Phi_{0}}{32\pi^{2}\lambda^{2}\xi}.\label{eq:jc-col}
\end{equation}
In anisotropic superconductors $\mathrm{j_{c}^{max}}$ is given in
\cite{Kulic-CD}. (In SI units $\mathrm{j_{c,SI}^{max}\approx\Phi_{0}/8\pi\mu_{0}\lambda^{2}\xi}$,
where $\lambda$$\mathrm{(T)}\approx\lambda_{0}(1-t)^{-1/2}$ and
$\xi(T)\approx\xi_{0}(1-t)^{-1/2}$ are the Ginzburg-Landau penetration
depth and coherence length, respectively). In $\mathrm{H_{3}S}$ one
has $\lambda\approx(1-2)\times10^{3}\textrm{Å}$ and $\xi_{0}\sim(15-20)\textrm{ Å}$
\cite{Drozdov-2015}, which gives for $\mathrm{j_{c0}^{max}}$ in
the range $\mathrm{j_{c0}^{max}}$$\approx(0.8-3)\times10^{8}\mathrm{A}/\mathrm{cm^{2}}$.
In the following, the critical current density $\mathrm{j_{c0}}$
will be estimated from experiments on the magnetic penetration in
$\mathrm{H_{3}S}$ \cite{Troyan}, where it is found $\mathrm{j_{c0}\approx(1,3-1,5)\times}10^{7}$$\mathrm{\mathrm{\mathrm{\textrm{Å}}/cm^{2}}}$,
which gives for the reducing factor $\mathrm{\eta\approx(0.1-0,3)}$. 

\subsection{Crossover field $\mathit{\mathrm{\mathit{B}_{rb}}}$ for the single
vortex to vortex bundles pinning }

The generic $\mathrm{H-T}$ phase diagram for the high temperature
superconductors with columnar pinning defects is shown in $\mathit{Fig.}$
$\mathit{1}$ - see \cite{Blatter-RMP}, and some magnetic properties
of $\mathrm{H_{3}S}$ will be discussed in the framework of this phase
diagram. In $\mathbf{II.B}$ the single-vortex pinning is considered,
which occurs at small magnetic field $\mathrm{B<B_{\phi}}$. When
the inter-vortex distance $\mathrm{a_{v}\approx(\Phi_{0}/B)^{1/2}}$
is larger than the average distance between columnar defects $\mathrm{d_{\phi}}=(\Phi_{0}/\mathrm{B_{\phi}})^{1/2}$
and the single vortex pinning energy is larger than the inter-vortex
energy, than the vortices accommodate freely to the pinning sites.
However, when $\mathrm{B>B_{\Phi}}$ the inter-vortex interaction
starts to be important for pinning and dynamical properties (such
as the vortex creep). In that case, vortex bundles are pinned. The
maximal crossover field $\mathrm{B_{rb}^{max}}$ - see $\mathit{Fig.}$
$\mathit{1}$, separates the single vortex regime from the vortex
bundle regime and it is obtained by comparing the energy of the elastic
shear deformation (of the order $\mathrm{u\sim d_{\Phi}}$) with the
maximal pinning potential $\mathrm{\epsilon_{p}^{max}}$ per unit
length, i.e. $\epsilon_{\mathrm{shear}}=\mathrm{c_{66}(d_{\phi}/a_{v})^{2}a_{v}^{2}}$$\backsimeq\mathrm{\epsilon}_{p}^{\mathrm{max}}$.
The shear modulus is given by $\mathrm{c_{66}\approx\mathrm{\mathrm{\varPhi_{0}B/(8\pi\lambda)^{2}}}}$what
gives for $\mathrm{B_{rb}}$ 
\begin{equation}
\mathrm{B_{rb}<B_{rb}^{max}\lesssim\frac{4\mathrm{\epsilon_{p}^{max}}}{\epsilon_{0}}B_{\phi}},
\end{equation}
where $\epsilon_{0}=\Phi_{0}^{2}/16\pi^{2}\lambda^{2}$ \cite{Blatter-RMP}.
Since $\mathrm{\epsilon_{p}^{max}=\epsilon_{0}/2}$ it gives, that
for $\mathrm{B<B_{br}<B_{rb}^{max}\approx2B_{\Phi}}$ the columnar
defects outnumber the vortices and the single-vortex pinning prevails.
Note, that the real crossover field is $\mathrm{B_{rb}<B_{rb}^{max}}$.
Since, $\mathrm{\eta_{col}<1}$ this inequality holds for $\mathrm{T\ll T_{c}}$,
where $\xi(T)\approx\xi_{0}$ and $\lambda(T)\approx\lambda_{0}$.
It is argued bellow, that in order to explain the $\mathrm{TBR}$
experiment in $\mathrm{H_{3}S}$ the regime $\mathrm{B>B_{rb}}$ is
also realized. 

\begin{figure}
\includegraphics[scale=0.4]{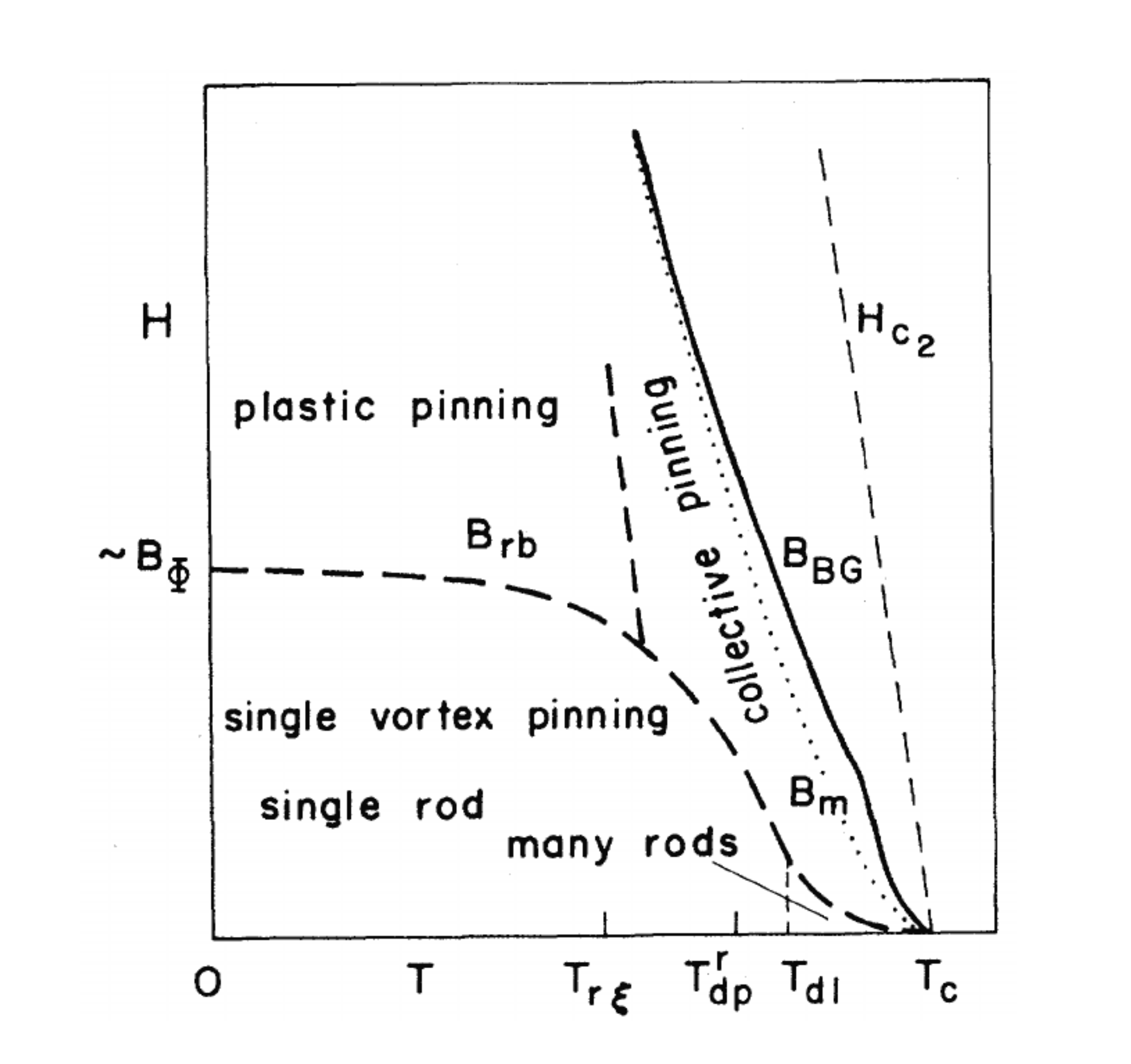}

\caption{The generic $H-T$ phase diagram for superconductors with $\mathit{columnar}$
$\mathit{defects}$. The doted melting line $\mathrm{B_{m}(T)}$ of
the pure sample is transformed into a Bose-glass transition line $\mathrm{B_{BG}(T)}$.
Also shown are the various pinning regimes with a single-vortex/single-rod
pinning region at low fields $\mathrm{\mu_{\mathrm{0}}H<B_{rb}(T)<2B_{\Phi}}$.
For $\mathrm{T<T_{dp}^{r}}$ the fluctuations of a vortex from one
defect to another are suppressed, while for $\mathrm{T>T_{dp}^{r}}$
the pinning potential is exponentially reduced. At $\mathrm{T>T_{dl}}$
the individual flux lines are pinned collectively by an assembly of
columnar defects (rods) at high temperatures. Above the crossover
line $\mathrm{B_{rb}(T)}$, the largest energy in the problem is the
inter-vortex interaction and pinning involves vortex bundles - taken
from \cite{Blatter-RMP}.}

\end{figure}

Note, that in cuprates with columnar defects, pinning properties are
highly anisotropic with the maximal critical current for the magnetic
field aligned along the columnar defect (and for $\mathrm{j_{c}\perp\mathbf{H}}$).
The latter is confirmed for irradiated $\mathrm{YBa_{2}Cu_{3}O_{7}}$
\cite{Civale}. It seems, that this is not the case for $\mathrm{H_{3}S}$,
where $\mathrm{j_{c0}^{\perp}\approx j_{c0}^{\parallel}}$ and similar
columnar densities of defects are realized along and perpendicular
to the sample surface, i. e. $\mathrm{B_{\phi}^{\perp}\simeq B_{\phi}^{\parallel}}$
\cite{Troyan}. This implies that the columnar defects in $\mathrm{H_{3}S}$
are oriented along, both, perpendicular and parallel, axes almost
equally - see $\mathit{Fig.}$ $\mathit{2}$. 

\subsection{Thermal vortex depinning from columnar defects}

For any high temperature superconductor thermal fluctuations of vortex
lines are important at higher temperatures, because of smoothing of
the pinning potential. This significantly lowers $\mathrm{j_{c}(T)}$
at and above some depinning temperature $\mathrm{T_{dp}^{r}}$, when
the effective pinning potential (per unit vortex length) $\epsilon_{p}(T)=\epsilon_{p}(0)\varphi(T)$
becomes small, since $\mathrm{\varphi(T>T_{dp}^{r})\ll1}$. For the
sake of clarity - there are two types of thermal motion of vortex
lines in the presence of pinning centers : ($\mathit{i}$) Phonon-like,
with small amplitude fluctuations affecting an individual pinning
potential - $\mathit{intravalley}$ fluctuations, thus smoothing the
pinning potential and reducing $\mathrm{j_{c}(T)}$ significantly
near $\mathrm{T_{dp}^{r}}$. ($\mathit{ii}$) The second kind of thermal
motion is related to large $\mathit{intervalley}$ thermal fluctuations,
which cause jumping of vortices from one to another pinning center
(valley). These are mainly responsible for the vortex-creep phenomena
- which is not studied here. Here, we deal with type ($\mathit{i}$)
thermal effects in the regime of the single-vortex pinning. Above
and at $\mathrm{T_{dp}^{r}}$ the amplitude of the thermal fluctuations
$\left\langle \mathrm{u}^{2}\right\rangle _{\mathrm{th}}^{1/2}$ increases
beyond the extent of the vortex core, $\left\langle \mathrm{u^{2}}\right\rangle _{\mathrm{th}}>\xi^{2}$.
In that case, the vortex experiences smaller averaged pinning potential.
The calculation of $\mathrm{T_{dp}^{r}}$ is sophisticated and based
either on: ($\mathit{i}$) the analogy of the vortex statistical physics
with columnar defects and the quantum 2D-Bose gas placed in a random
pinning potential, or ($\mathit{i\mathit{i}}$) on the statistical
physics of vortices \cite{Blatter-RMP}. It turns out that for $\mathrm{r\gtrsim\sqrt{2}\xi_{0}}$
the depinning temperature $\mathrm{T_{dp}^{r}}$ is approximately
given by the self-consistent equation $\mathrm{T_{dp}^{r}\approx r\cdot\sqrt{\epsilon_{p}(T_{dp})\epsilon_{0}(T_{dp})}}$,
where $\mathrm{\epsilon}_{p}(T)=\eta_{\mathrm{col}}\cdot\Phi_{0}^{2}/32\pi^{2}\mathrm{\lambda^{2}(T)}$
and $\epsilon_{0}=\Phi_{0}/16\pi^{2}\mathrm{\lambda^{2}}(T)$ \cite{Blatter-RMP}.

\section{Temperature broadening of resistance in the magnetic field in $\mathrm{H_{3}S}$}

In Refs. \cite{Hirsch-1-1}-\cite{Hirsch-Marsiglio-arXiv V5-2022-1}
it was claimed that in order to explain the temperature broadening
of the resistance in magnetic field ($\mathit{TBR}$), $\delta t_{c}(h)\equiv(T_{c}-T_{c}(h))/T_{c}$,
in HP-hydrides and in the framework of physics of soft superconductors,
it is necessary to invoke an unphysically large critical current density
$\mathrm{j_{c0}>10^{9}}$$\mathrm{A/cm^{2}.}$ In the case of the
questionable CSH superconductor even much larger critical current
is needed, i.e. $\mathrm{j_{c0}>10^{11}}$$\mathrm{A/cm^{2}}$- where
$\delta t_{c}(h)$ is field independent \cite{Snider}. In the following
it is argued, that the reduction of $\delta t_{c}(h)$ in $\mathrm{H_{3}S}$
can be explained by invoking the SCPC model - which assumes that $\mathrm{H_{3}S}$
is a $\mathit{hard}$ $\mathit{type-II}$ $\mathit{superconductor}$
with elongated intrinsic columnar pinning defects. The geometry of
the experiment is shown in $\mathit{Fig.}$ $\mathit{2}$. 

Let us briefly introduce the reader into the subject of $\mathit{TBR}$
in $\mathrm{H_{3}S}$, which is based on the Tinkham theory for $\mathit{TBR}$
\cite{Tinkham-1} and the SCPC model \cite{Kulic-RTSC-v1}. 

\begin{figure}
\includegraphics[scale=0.3]{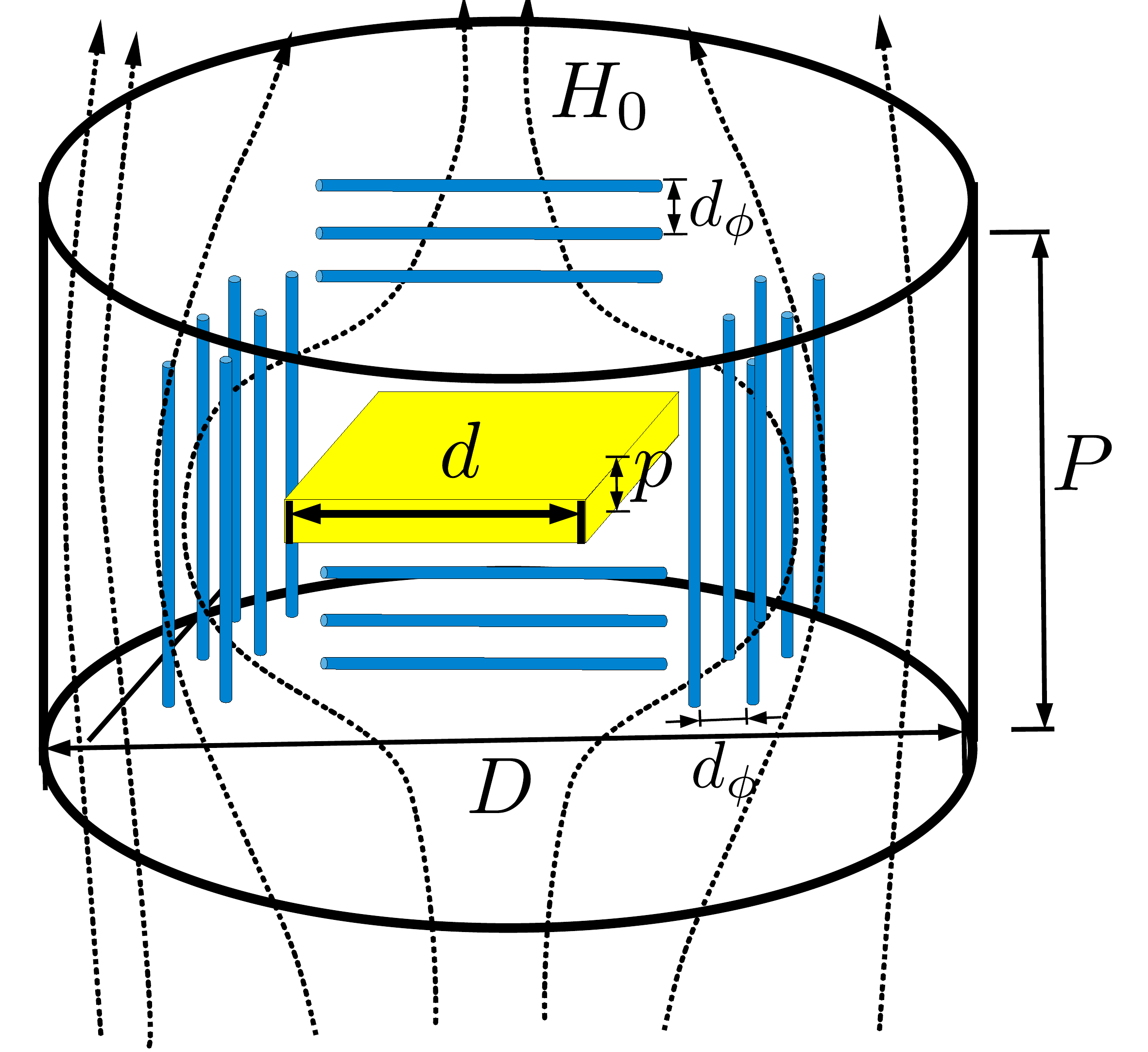}\caption{Schematic view of the experiment of the magnetic flux trapping and
penetration in the $H_{3}S$ disk-like sample \cite{Kulic-RTSC-v1}:
$\mathit{D=30\mu m;}P=5\mu m$ for the perpendicular geometry $\mathrm{\mathbf{H}_{0}\perp D}$.
According to the SCPC-model, long columnar defects ($\mathit{blue}$
$\mathit{cylinders}$) strongly pin and trap vortices making huge
magnetization hysteresis and the critical current density $\mathit{j_{c}\sim\delta M}$.
The non-superconducting $^{119}$Sn film ($\mathit{yelow}$ with $\mathit{d=20\mu m;\textrm{ }}p=2.6$$\mu m$)
is implemented in the experiment for the detection of the penetrated
magnetic field \cite{Troyan}. Similar analyzes holds for the parallel
geometry $\mathrm{\mathbf{H}_{0}\parallel D}$.}
\end{figure}
Namely, in all superconductors dissipationless current can flow in
the vortex state with pinning defects. However, when the pinning energy
is small, especially for $\mathrm{T}$ near $\mathrm{T_{c}}$, vortices
jump easily from one center to another under temperature fluctuations,
thus giving rise to a vortex motion and dissipation of energy - called
flux creep \cite{Anderson}. These jumps are activation-like and proportional
to the escape probability (from the pinning center) $\mathit{exp\mathrm{(-U_{p}/T)}}$
. Since for point defects $\mathit{\mathrm{U}_{\mathrm{p}}^{\mathrm{pd}}}$$\sim\xi^{3}$,
then this energy barrier is small in superconductors with small $\xi$,
what is, for instance, the origin of a pronounced dissipation in HTSC-cuprates
(with oxygen vacancies). In that sense, the long pinning defects with
the energy $U$$_{p}$ $\sim L_{v}\xi^{2}$ make this barrier much
higher, thus suppressing the dissipation effects significantly. In
magnetic fields much higher than the lower critical field, i.e. for
$\mathit{\mathrm{H_{c1}\ll H<H_{c2}}}$, one has $\mathit{\mathrm{B\approx\mu_{0}H}}$
and the vortex distance is given by $\mathrm{\mathit{\mathrm{a}\approx(\mathbf{\mathrm{\Phi_{0}/B}})^{1/2}}<\lambda}$.
In that case bundles of vortices, each with the surface $\sim a^{2}$,
are pinned \cite{Yeshurun} with the pinning energy of the bundle
$\mathit{\mathrm{U_{p}}}$$\sim\mathrm{L_{v}}a^{2}$. The Tinkham
$\mathit{TBR}$ theory \cite{Tinkham-1} applied to such problems
is based on the Ambegaokar and Halperin theory for thermally activated
phase motion in Josephson junctions \cite{Ambegaokar}. As the result,
it gives for the resistance of the superconductor (with small transport
current) \cite{Tinkham-1}

\begin{equation}
R/R_{N}\approx\left[I_{0}(\gamma/2)\right]^{-2},\gamma=U_{p}/T,\label{eq:R-Tink}
\end{equation}
where $\mathrm{R_{N}}$ is the resistance of the normal state at $\mathit{T}_{c}$
and $\mathit{I_{0}}$ is the modified Bessel function. 

In the $\mathrm{SCPC}$ model with $\mathit{columnar}$ $\mathit{pinning}$
centers with $\mathit{\mathrm{L\approx L_{v}}}$ and near $\mathit{T_{c}}$
one obtains for $\gamma$

\begin{equation}
\gamma^{\mathrm{scpc}}=\beta_{K}\left(\frac{L_{v}}{\xi_{0}}\right)\frac{(1-t)^{2}}{h}(2\pi\xi_{0}^{2}\frac{j_{c0}\Phi_{0}}{cT_{c}}),\label{eq:gamma-0}
\end{equation}
where $\mathit{\mathrm{h}\approx\mathrm{H/H_{c2}(0)}},$$\mathrm{\delta t_{c}\equiv(1-t)\equiv1-T/T_{c}}$
and $\beta_{K}\approx1$. In the case of $\mathit{point}$ $\mathit{defects}$
one has for $\gamma$
\begin{equation}
\gamma^{\mathrm{pd}}\approx\frac{(1-t)^{3/2}}{h}(2\pi\xi_{0}^{2}\frac{j_{c0}^{\mathrm{pd}}\Phi_{0}}{cT_{c}}).\label{eq:gamma-pd}
\end{equation}
Here, the critical current for weak pinning $\mathrm{j_{c0}^{pd}=\eta_{pd}\cdot j_{c0}}$
is defined via the weak pining energy $\mathrm{U_{p}^{pd}\approx\eta_{pd}H_{c}^{2}\xi^{3}}$
with $\eta_{\mathrm{pd}}<1$ , while in the SCPC model one has $\mathrm{j_{c0}=\eta_{col}\cdot j_{co}^{max}}$
- see $\mathrm{Eq.}$ $\mathrm{(1)}$. The critical current density
for weak pinning centers is usually of the order $\mathrm{j_{c0}^{pd}\sim10^{6}}$
$\mathrm{A/cm^{2}}$. Note, that $\gamma^{\mathrm{scpc}}\sim(\mathrm{1-t})^{2}/h$
for long columnar defects, while $\gamma^{pd}\sim(\mathrm{1-t})^{3/2}/h$
for point defects. The latter fact gives rise to the different temperature
dependence of the irreversible line $\mathrm{H_{irr}(T)}$ in HTSC
(with oxygen vacancies as point defects) and $\mathrm{H_{3}S}$ -
see more below. From these expressions it is seen, that the relative
barrier in the case of long columnar pinning centers is larger by
the factor $\mathit{\mathrm{(L_{v}/\xi_{0})\gg1}}$, compared with
the one for point-like randomly distributed defects - used in \cite{Hirsch-1-1}-\cite{Hirsch-Marsiglio-arXiv V5-2022-1}.
This means that $\mathrm{j}_{c0}^{\mathrm{pd}}$ for point defects
is replaced by much larger quantity $\mathit{\mathrm{(L_{v}/\xi_{0})\times j_{c0}}}$
in the SCPC-model - where one has $\mathit{\mathrm{(L_{v}/\xi_{0})\times j_{c0}}}$$\gg$$\mathrm{j_{c0}}$. 

Let us compare the prediction of the standard theory for weak pinning
(with point-like defects) applied to $\mathrm{H_{3}S}$ - done in
\cite{Hirsch-1-1}-\cite{Hirsch-Marsiglio-arXiv V5-2022-1}. In the
case when the resistance is measured at the 10 \% level, i.e. $\mathit{R/R_{N}=\mathrm{0.1}}$,
it gives for $\mathit{I_{0}\mathrm{(\gamma/2)}\approx\mathrm{3.2}}$
and $\gamma\approx5(=\gamma^{scpc}=\gamma^{pd}$ in both cases. In
the field $\mathit{\mathrm{B\approx\mu_{0}H\approx1}}\mathrm{T}$
and for $\mathit{\mu_{0}\mathrm{H_{c2}\mathrm{(0)}}\approx\mathrm{100}}$
$\mathrm{T}$ one has $\mathrm{\mathrm{\mathit{\mathrm{h}\approx\mathrm{0.01}}}}$.
We assume also the following realistic parameters for the $\mathrm{HP}$-hydrides:
$\xi_{0}\approx20\mathrm{\textrm{ Å}}$ and $\mathit{\mathrm{T}_{c}}$$\sim200$
$\mathrm{K}$, $\Phi_{0}=2\times10^{-7}G\times cm^{2}$. If one takes
$\mathrm{j_{c0}^{pd}=}$$\alpha\mathrm{j}_{c0}\approx\alpha\times10^{7}$
$\mathrm{A/cm^{2}}$ with $\alpha\sim0.1$, then by using Eq.(\ref{eq:gamma-pd})
one obtains
\begin{equation}
\delta t_{c}^{pd}\equiv\frac{\delta T_{c}^{pd}(h=0.01)}{T_{c}}\approx\frac{0.1}{\alpha}.\label{eq:tc-pd}
\end{equation}
This is a too large value ($\delta t_{c}^{pd}>0.1$) - see the red
line in $\mathit{Fig.}$ $\mathit{3}$, compared to the experimental
one $\delta t_{c}^{\mathrm{exp}}\apprle10^{-2}$ \cite{Eremets-Meissner-2022}.
Based on Eq.(\ref{eq:gamma-pd}) one concludes that in order to explain
the experimental value $\delta t_{c}^{exp}(h=0.01)\apprle10^{-2}$
in $\mathrm{H_{3}S}$ in the framework of the weak pinning theory
one needs a mach larger (effective) critical current $\mathrm{j_{c0}^{pd}}>3\cdot$$10^{8}$
$\mathrm{A/cm^{2}}$. However, the latter value is far beyond the
range of the weak pinning theory.

However, the experimental value $\delta t_{c}^{\mathrm{exp}}\apprle10^{-2}$
can be explained by the SCPC model with the long columnar pinning
defects, where the $\delta t_{c}^{\mathrm{scpc}}(h)$ depends on the
large factor $\mathrm{(L/\xi_{0})\times j_{c0}}$, thus making $\delta\mathrm{t}_{c}^{\mathrm{scpc}}(h)$
small. In the case of $\mathrm{H_{3}S}$ where $\gamma_{00}=2\pi$$\xi_{0}^{2}$$\mathrm{\times j_{co}}$$\Phi_{0}/cT_{c}$
and $\mathrm{j_{c0}}$ is in the range $\mathit{\mathrm{j_{c0}}}>10^{7}$$A/cm^{2}$
(in CGS units $\mathrm{\mathit{\mathrm{j}_{\mathrm{\mathrm{c}o}}}}$$>3\times10^{16}\textrm{ }\mathrm{esu/cm^{2}}\textrm{)}$
and for $\mathit{\mathrm{L}_{v}\sim\mathrm{(0.5-1)\textrm{ }}\mu m}$
one obtains
\begin{equation}
\delta t_{c}^{\mathrm{scpc}}(h)\approx\left(\frac{5\xi_{0}h}{L_{v}\gamma_{00}}\right)^{1/2}\lesssim0.01.\label{eq:Ratio}
\end{equation}
The obtained result for $\mathit{TBR}$ in $\mathrm{Eq.}$ $\mathrm{(7)}$
in the SCPC-model for $\mathrm{h\approx0.01}$ and $\mathrm{j_{c0}\sim10^{7}}$$\mathrm{A/cm^{2}}$
is in satisfactory agreement with the experimental values shown in
$\mathit{Fig.}$ $\mathit{3}$ - the blue line. Since $\delta t_{c}^{\mathrm{scpc}}\ll$$\delta t_{c}^{pd}$
this means that the SCPC model is able to describe $\mathit{TBR}$
in the $\mathrm{H_{3}S}$ superconductor. 

\begin{figure}
\includegraphics[scale=0.45]{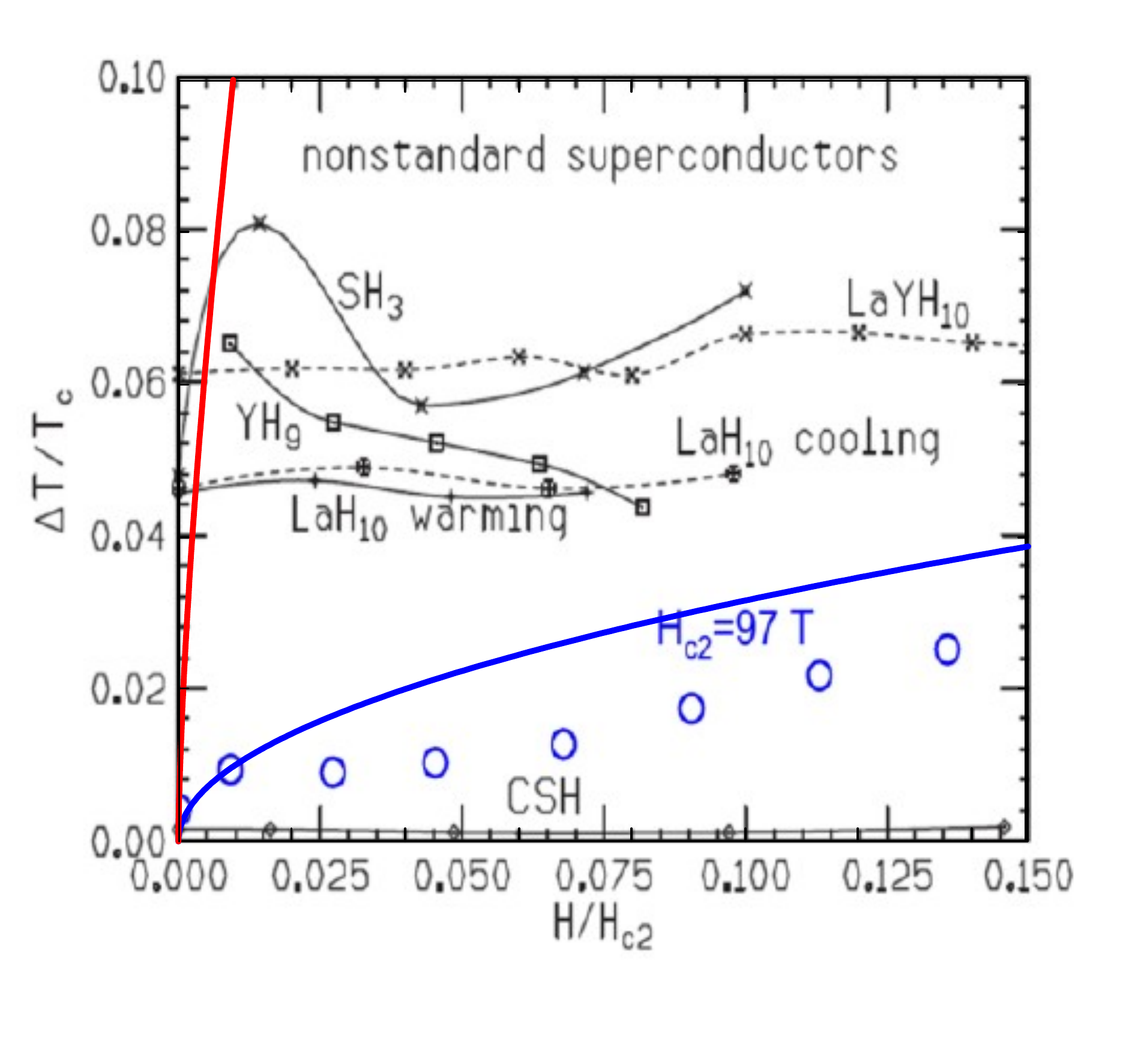}

\caption{Broadening of the superconducting transition ($\mathit{TBR}$) $\delta t_{c}(h)\equiv\Delta T/T_{c}$
under external magnetic fields in different superconducting HP-hydrides
derived in Ref. \cite{Hirsch-1-1} - points and solid lines and the
experimental values for TBR in $\mathrm{H_{3}S}$ extracted in \cite{Eremets-Meissner-2022}
- $\mathit{blue}$ $\textrm{\ensuremath{\mathit{points}}}$. $\mathit{Blue}$
$\mathit{line}$ - the theoretical line predicted by the SCPC-model
$\delta\mathrm{t^{scpc}}_{c}(h)\sim h^{1/2}$ ($\mathrm{h=H/H_{c2}}$,
$\mathrm{H_{c2}\approx100}$$\mathrm{T}$) for $\mathit{j_{c0}}\approx10^{7}$$A/cm^{2}$
and $\mathrm{L\approx0.5}$$\mu m$. The model is more suitable for
low $\mathrm{h\protect\leq0.01}$- see text. $\mathit{Red}$ $\mathit{line}$
- the prediction of the standard model for $\mathit{TBR}$ with weak-pinning
$\delta t_{c}^{pd}(h)\sim h^{2/3}$ and for $\mathit{j_{c0}}\approx10^{7}$$A/cm^{2}$
is inadequate for $\mathrm{H_{3}S}$ .}

\end{figure}

Note, that there is $\mathit{TBR}$ also in the zero magnetic filed
($\mathrm{H=0}$), i.e. there is an intrinsic $\mathit{TBR}$, $\delta T_{0}\neq0$
, which should be taken into account in the analyzes of experiments.
This was done in \cite{Eremets-Meissner-2022} and \cite{Hirsch-1-1},
where one has $\delta T_{c}(h)$$\approx$$\delta T_{\mathrm{tot}}(h)-\delta T_{c0}$
and $\delta T_{\mathrm{tot}}$ is the total $\mathit{TBR}$. The extracted
experimental results for $\delta t_{c}(h)(\equiv\delta T_{c}(h)/T_{c})$
in $\mathrm{H_{3}S}$ are shown in $\mathit{Fig.}$ $\mathit{3}$
\cite{Eremets-Meissner-2022} - blue circles, for $\mathrm{h}$ within
the range of $\mathrm{0<h<0.15}$.

We stress again, that in order to explain the much smaller $\mathit{TBR}$
, $\delta t_{c}(h)$, in $\mathrm{H_{3}S}$ (and in other HP-hydrides)
than the standard Tinkham theory predicts for the weak pinning - see
Eq.(\ref{eq:gamma-pd}), the authors of \cite{Hirsch-1-1}-\cite{Hirsch-Marsiglio-arXiv V5-2022-1}
assume an unrealistically large critical current density $\mathrm{\mathit{\mathrm{j}^{\mathrm{pd}}}}$$\sim(10^{9}-10^{11})\mathrm{A/\mathrm{cm^{2}}}$.
The latter value is much larger than the experimental one $\mathit{\mathrm{j_{c0}^{exp}}}$$\apprge10^{7}A/\mathrm{cm^{2}}$.
The second possibility is to call into question the existence of superconductivity
in HP-hydrides. This (second) possibility is accepted in \cite{Hirsch-1-1}-\cite{Hirsch-Marsiglio-arXiv V5-2022-1}.
However, in the proposed SCPC model $\mathrm{j_{c0}}$ in the formula
for $\delta t_{c}(h)$ is in fact replaced by the much larger quantity
$\mathit{\mathrm{(L_{v}/\xi_{0})\times j_{c0}}}$ - see Eq.(\ref{eq:gamma-0}),
which for $\mathit{\mathrm{(L_{v}/\xi_{0})\sim10^{3}}}$ gives realistic
values for $\mathrm{\mathrm{\mathrm{\mathit{\mathrm{j}_{c\mathrm{0}}}}\apprge10^{7}}}$
$\mathrm{A/cm^{2}}$. This analysis confirms our claim, that there
is no reason to call into question the existence of the conventional
superconductivity in $\mathrm{H_{3}S}$.

\section{penetration of the magnetic field in $\mathrm{H_{3}S}$}

In order to confirm the existence of superconductivity in a sample,
two kinds of experiments in an external magnetic field are necessary:
(i) At $\mathrm{T>T_{c}}$ the sample is placed in a magnetic field
$\mathbf{\mathrm{H<H_{c1}}}$(or $\mathrm{H_{c}}$ in type-I superconductors)
and than cooled below $\mathrm{T_{c}}$ - the $\mathit{FC}$ (field
cooled) experiment. If the magnetic field is $\mathit{expelled}$
from the sample the $\mathit{Meissner}$ $\mathit{effect}$ is realized.
Its existence means a definitive proof for superconductivity in the
sample; (ii) In the $\mathit{zero}$ $\mathit{field}$ $\mathit{cooled}$
($\mathit{ZFC}$) experiment the sample is first cooled to $\mathrm{T<}\mathrm{T_{c}}$
at zero external field ($\mathrm{H=0}$) and then is the field $\mathrm{H<H_{c1}}$
applied. In this case, the magnetic field is $\mathit{excluded}$
from the sample - the so called $\mathit{diamagnetic}$ $\mathit{shielding}$
$\mathit{mode}$. The latter effect is in fact due to the well known
Lenz law in electrodynamics. In the Meissner state of type-I superconductors
currents in a bulk sample decay exponentially from the sample surface
and the average current over the bulk sample is zero. In an ideal
type-II superconductor with regular vortex lattice the average circulating
current (around the vortices) is also zero. A net average (transport)
current can exist only due to distortions of the vortex lattice in
the presence of pinning defects. As a result, the transport current
can flow in bulk type-II superconductors, $j(r)\neq0$ , and the field
can penetrate into the sample at much larger distance than the penetration
depth $\lambda$. If the pinning is strong (and the critical current
large) such superconductors are called hard superconductors. 

In Section III the theoretical (in the SCPC-model) and experimental
results for $\mathit{TBR}$ suggest, that $\mathrm{H_{3}S}$ (and
other HP-hydrides) is a hard type-II superconductor, with large Ginzburg-Landau
parameter $\kappa(=\lambda/\xi)=(50-100)$ and high (low-temperature)
critical current density $\mathit{\mathrm{j}_{c\mathrm{0}}}$$\apprge10^{7}A/\mathrm{cm^{2}}$.
Next questions related to $\mathit{\mathrm{H_{3}S}}$ are: ($\mathit{i}$)
What is the prediction of the SCPC model for the penetration of magnetic
field ($\mathit{PMF}$) into a superconducting sample; ($\mathit{ii}$)
How big is the magnetic hysteresis observed in $\mathrm{H_{3}S}$
\cite{Troyan}? For simplicity, these questions are analyzed quantitatively
in the framework of the Bean critical state model - first for a long
cylinder and than for a disk with small thickness $\mathrm{P(\ll D)}$.
Afterwards, the theory is compared with the experiment in $\mathrm{H_{3}S}$
\cite{Troyan}. In the following, the results are presented either
in the CGS or in SI units - according to the existing magnetic measurements.
$\mathrm{B^{0}}$, $\mathrm{B_{0}}$ and $\mathrm{B_{ext}}$ are equivalent
labeling for the applied magnetic field (induction). Some problems
related to the $\mathit{FC}$ Meissner effect are discussed at the
end of this Section. 

\subsection{Experimental results for the ZFC field penetration in $\mathrm{H_{3}S}$}

The $\mathit{ZFC}$ penetration depth is studied experimentally in
$\mathrm{\mathrm{H_{3}}S}$ by the novel NRS technique \cite{Troyan}
- the geometry of the ZFC experiment is shown in $\mathit{Fig.}$
$\mathit{2}$. The experiment measures the time evolution of the scattered
radiation re-emitted by the $^{119}$Sn film - $\mathit{Fig.}$ $\mathit{3}$
in Ref. \cite{Troyan}. In this ZFC experiment a non-superconducting
thin $^{119}$Sn film is placed inside the $\mathrm{H_{3}S}$ disk-sample
(see $\mathit{Fig.}$ $\mathit{2}$ - yellow color) being used as
a sensor of the magnetic field. The internal magnetic field on the
$^{119}$Sn sensor is monitored by the nuclear resonance scattering
of synchrotron radiation. When the magnetic field penetrates in the
$^{119}$Sn non-superconducting film, the experimental curve for the
radiation intensity as a function of time shows $\mathit{quantum}$
$\mathit{beats}$, due to the interference of the split $^{119}$Sn
nuclear levels by the magnetic field. These beats are indeed observed
in the field $\mathrm{B^{0}=0.68}$ $\mathrm{T}$ at $\mathrm{T>T_{c}}$
- when the sample is non-superconducting and the perpendicular magnetic
field penetrates completely into the $^{119}$Sn film - $\mathit{Fig.}$
$\mathit{3}$ in Ref. \cite{Troyan}. Note, that since $\mathrm{H_{3}S}$
is a type II superconductor one expects that if the sample is free
of pinning centers, then for $\mathrm{B^{0}\gg\mu_{0}H_{c1}=}(18-60)$
$\mathrm{mT}$ the magnetic field should enter into the whole sample
in the form of (almost) homogeneously distributed vortices - thus
showing quantum beats in the $^{119}$Sn film. However, in the $\mathit{perpendicular}$
$\mathit{geometry}$ of the experiment ($\mathrm{\mathbf{H}\perp D}$
in $\mathit{Fig.}$ $\mathit{2}$) at low temperatures and in the
field $\mathrm{B^{0}=0.68}$ T \cite{Troyan} the vortices do not
show up in the center of the $^{119}$Sn film up to some temperature,
i.e. there are no quantum beats - $\mathit{Fig.}$ $\mathit{3}$ in
\cite{Troyan}. This means that the vortices are strongly pinned in
the region outside of the $^{119}$Sn film. Strong pinning of vortices
in a superconductor implies, that the magnetization of the sample
is strongly hysteretic, as revealed experimentally in $\mathrm{H_{3}S}$
- see \cite{Eremets-Meissner-2022},\cite{Drozdov-2015},\cite{Minkov-Meissner-2021}.

\begin{figure}
\includegraphics[scale=0.4]{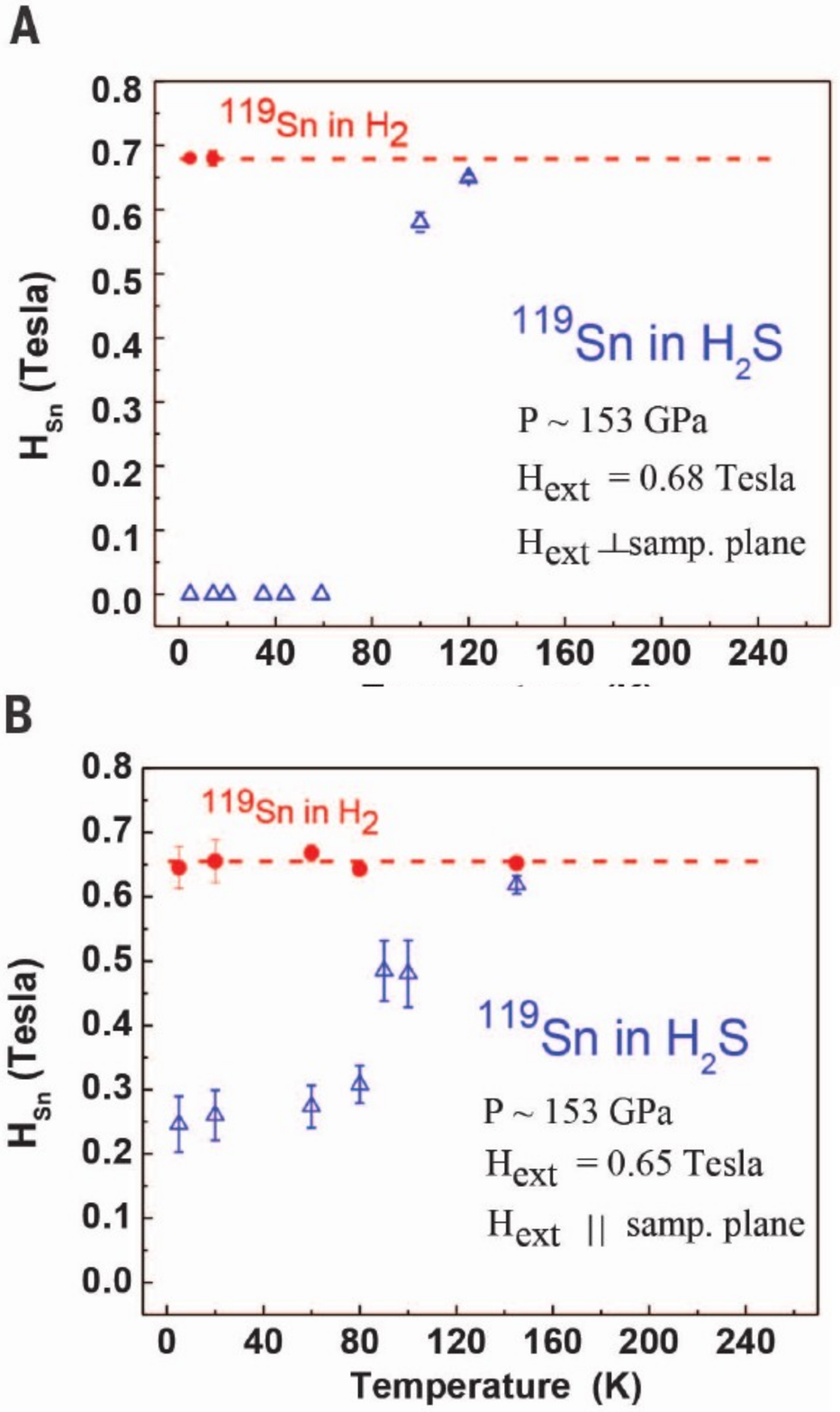}

\caption{The experimental temperature dependence of the magnetic field on the
sensor $^{119}$Sn film (placed) inside $\mathrm{H_{3}S}$ (see $\mathit{Fig.}$
$\mathit{2}$) at 153 GPa shown by $\mathit{blue}$ $\mathit{triangles}$.
The external field at the reference non-superconducting sample $^{119}$Sn
in $\mathrm{H_{2}}$ at 150 GPa is shown by $\mathit{red}$ $\mathit{dots}$
- see \cite{Troyan}. ($\mathbf{A\mathrm{)}}$ and $\mathbf{\mathrm{(}B}$)
are measurements in the perpendicular ($\mathrm{H_{ext}^{\perp}\perp D}$)
and parallel ($\mathrm{H_{ext}^{\parallel}\parallel D}$) geometry
of the external magnetic field, respectively. Dashed lines are guides
to the eye. Taken from \cite{Troyan}.}
\end{figure}

Two geometries were studied in \cite{Troyan}: ($\mathbf{\mathrm{\mathit{A}}}$)
The $\mathit{perpendicular}$ $\mathit{geometry}$ - where the field
is perpendicular to the disk surface ($\mathrm{\mathbf{H}\perp D}$
in $\mathit{Fig.}$ $\mathit{2}$); ($\mathit{\mathbf{\mathrm{B}})}$
The $\mathit{parallel}$ $\mathit{geometry}$ - the field is parallel
to the disk surface ($\mathrm{\mathbf{H}\parallel D}$). As it is
seen in $\mathit{Fig.}$ $\mathit{4A}$ the average field inside the
$^{119}$Sn sensor in the perpendicular geometry is approximately
zero for $\mathrm{T<80}$ $\mathrm{K}$ and reaches the value $\mathrm{B_{ext}=}$$0.68$
$\mathrm{T}$ at around $\mathrm{T}$$\approx120$ $\mathrm{K}$.
In the parallel geometry the field inside the $^{119}$Sn film is
finite for $\mathrm{T>0}$ (but smaller than $\mathrm{B_{ext}=}$$0.68$
$\mathrm{T}$) starting to increase at $\mathrm{T>80}$ $\mathrm{K}$
saturating to $\mathrm{B_{ext}=}$$0.68$ $\mathrm{T}$ at $\mathrm{T}$$\gtrsim140$
$\mathrm{K}$ - $\mathit{Fig.}$ $\mathit{4B}$. In the next Subsection
these experimental results are explained by the theory which is based
on the Bean critical state model and the SCPC model with high critical
current density $\mathrm{j_{c0}\apprge10^{7}}$$\mathrm{A/cm^{2}}$. 

\subsection{Penetration of the magnetic field in a long cylinder with pinning
centers}

In Section III it is argued, that the experimental results on the
thermal broadening of resistance ($\mathit{TBR}$) in magnetic filed
in $\mathrm{H_{3}S}$, $\delta t_{c}^{\mathrm{exp}}(h)$, give much
smaller value than those predicted by the Tinkham theory for standard
superconductors with weak pinning, i. e. $\delta t_{c}^{\mathrm{exp}}(h)$$\ll$
$\delta t_{c}^{pd}(h)$. However, the SCPC model with the columnar
pinning centers and large critical currents $\mathrm{j_{c}(0)}>10^{7}$$\mathrm{A/cm^{2}}$
explains this property, i. e. $\mathrm{\delta t_{c}^{scpc}(h)\approx\delta t_{c}^{exp}(h)}.$
Note, that the large value of $\mathrm{j_{c}(0)}$ inevitably causes
large magnetization hysteresis, which is seen experimentally in $\mathrm{H_{3}S}$
\cite{Eremets-Meissner-2022},\cite{Drozdov-2015},\cite{Minkov-Meissner-2021}.
The critical current density $\mathrm{j_{c}(0)}$ is a measure of
the strength of pinning forces. Below, it is shown that the large
value of $\mathrm{j_{c}(0)}$ is compatible with experiments for the
penetration of the magnetic field into thin $\mathrm{H_{3}S}$ film
\cite{Troyan}. To maximally simplify the problem a long cylinder
placed in the external field $\mathrm{B_{0}=\mu_{0}H_{0}}$ is first
considered (in absence of external transport currents), thus escaping
demagnetization effects. This case is also used in studying the field
penetration in the parallel geometry of $\mathrm{H_{3}S}$, i.e. when
$\mathrm{\mathbf{B}_{0}\parallel D}$.

In an homogeneous system in thermodynamic equilibrium, in an ideal
type II superconductor (without pinning centers), vortices are homogeneously
distributed over the bulk sample. In that case the macroscopically
averaged (over the sample) local magnetic induction is constant, i.e.
$\mathbf{\mathbf{\mathrm{\mathbf{\overline{B}}_{\mathrm{eq}}}}}$$(\mathbf{r}$)$\mathrm{=const}$
and the macroscopic local magnetization current density (averaged
over the vortex unit cell) is zero, i.e. $\mathrm{\mu_{0}\mathbf{\bar{j}_{\mathrm{eq}}}\mathrm{(\mathbf{r})=rot}\overline{\mathbf{B}}_{eq}}$=$\mathrm{0,}$
as well as the Lorentz force per unit volume of the vortex lattice
$\mathbf{f^{\mathrm{eq}}\mathrm{_{L}}=\mathrm{\mathbf{\bar{j}_{\mathrm{eq}}}}}\times\mathbf{\overline{B}_{\mathrm{eq}}}$$\mathrm{=0}$.
However, in the presence of pinning centers $\overline{\mathbf{B}}$$\mathrm{(\mathbf{r})}$
is $\mathit{inhomogeneous}$ and $\bar{\mathrm{\mathbf{j}}}$$\mathrm{(\mathbf{r})}$$\neq0$,
thus producing finite Lorentz force (per unit volume) on vortices
$\mathbf{f\mathrm{_{L}}=\mathrm{\mathbf{\bar{j}}}}\times\mathbf{\overline{B}}$$\neq0$.
When the vortices are pinned there is a pinning force (per unit volume)
$\mathbf{f}$$_{p}$ which counteracts the Lorentz force. In the static
case the vortices are not moving and the condition $\mathbf{f}$$_{L}=-\mathbf{f}_{p}$
is fulfilled everywhere in the sample. So, by increasing the applied
field the vortices are so rearranged that locally the maximal critical
current $\mu_{0}\mathrm{\mathbf{j_{\mathrm{c}}}}$$(B)=\mathrm{rot}\mathbf{\overline{B}}$
is achieved. 

The simplest, but very useful, model for the critical state is $\mathit{the}$
$\mathit{Bean}$ $\mathit{critical}$ $\mathit{state}$ $\mathit{model}$
\cite{Bean-1}, which assumes that $\mathrm{j_{c}}$ $\mathit{is}$
$\mathit{independent}$ $\mathit{on}$ $\mathrm{B}$. In the case
of a $\mathit{long}$ $\mathit{cylinder}$ $L$$\gg R$ (no demagnetization
effects) one has $\mathrm{d\overline{B}/dr}=\pm\mu_{0}\overline{j}_{c}$
and $\mathrm{\overline{B}(r)}$ is given by
\begin{equation}
B(r,B_{0})=B_{0}-B^{*}(T)(1-\frac{r}{R}).\label{eq:field}
\end{equation}

It is seen from Eq.(\ref{eq:field}) that for an applied filed $\mathrm{0\leq B_{0}<B^{*}}$
the field $\mathrm{B(r)}$ penetrates up to the point $\mathrm{r_{B_{0}}}=(1-\mathrm{B}_{0}/\mathrm{B^{*}}$$)R$,
where $\mathrm{B(r_{B_{0}},B_{0})=0}$. At $\mathrm{B_{0}=B^{*}}$
$\mathit{the}$ $\mathit{field}$ $\mathit{reaches}$ $\mathit{the}$
$\mathit{center}$ $\mathit{of}$ $\mathit{the}$ $\mathit{cylinder}$,
i.e. at $\mathrm{r_{B^{*}}=0}$ one has $\mathrm{B(r_{B^{*}}=0,B^{*})=0}$
- see $\mathit{Fig.}$ $\mathit{5}$. In the experiment \cite{Troyan}
- schematically given in $\mathit{Fig.}$ $\mathit{2}$, one has $\mathrm{D=30}$
$\mathrm{\mu m}$ and by assuming that $\mathrm{L\gg D}$ one has
$1.8$ $\mathrm{\mathrm{T<}B^{*}(j_{c0})}$$\mathrm{<18}$ $\mathrm{T}$.
This range for $\mathrm{B^{*}(j_{co})}$ is due to $\mathrm{j_{c0}}$
in the range $\mathrm{j_{c0}}\approx(10^{7}-10^{8})$ $\mathrm{A/cm^{2}}$. 

However, the applied field in the experiment by Troyan et al. \cite{Troyan}
was fixed to $\mathrm{B_{0}=0.68}$ $\mathrm{T}$ , i.e. $\mathrm{B_{0}}$$\mathrm{<B^{*}(T\ll T_{c})}$.
This means that at very low temperature ($\mathrm{T}$$\ll T_{c}$)
the field does not penetrate into the center of the non-superconducting
$^{119}$$\mathrm{Sn}$ film, if the sample would be a long cylinder
with $\mathrm{L(=P)\gg D}$. Note, that the magnetic field reaches
the front of the $^{119}$S film at $\mathrm{r_{B_{0}^{S}}=10}$$\mu m$,
i.e. for $\mathrm{B_{0}^{S}=(1/3)B^{*}}$ where $\mathrm{B_{0}^{S}}$$\approx(0.6-6$)
$\mathrm{\mathrm{T}}$, for $\mathrm{j_{c0}\approx(10^{7}-}10^{8}$$\mathrm{A/cm^{2}}$.
Therefore, in the case of a long cylinder with $\mathrm{L(=P)\gg D}$
at low $\mathrm{T}$ with $\mathrm{j_{c0}\approx1.5\times10^{7}}A/\mathrm{cm^{2}}$
placed in the external field $\mathrm{B_{0}=0.68}$ $\mathrm{T}$
, the field would not be able to penetrate neither the front of the
$^{119}$S film (where $\mathrm{r_{B_{0}^{S}}=10}$$\mathrm{\mu m}$)
nor the center ($\mathrm{r_{B^{*}}=0}$) of the sample, since $\mathrm{B_{0}<B_{0}^{S}}<\mathrm{B^{*}}$. 

We stress again, that the experiment in Ref. \cite{Troyan} was performed
on a finite-size sample in two geometries: ($\mathrm{A}$) the $\mathit{perpendicular}$
geometry when the magnetic field is perpendicular to the thin cylinder,
i.e. $\mathrm{\mathbf{B}_{0}\perp D}$ - see $\mathit{Fig.}$ $\mathit{2}$,
and ($\mathrm{B}$) in the $\mathit{parallel}$ $\mathit{geometry}$
when the field is parallel to the diameter $\mathrm{D}$ of the sample,
i.e. to the surface of the thin disk with $\mathrm{P\ll D.}$ In the
next Subsection it is shown, that in the case ($\mathrm{A}$) finite
size effects are very important, while in the case ($\mathrm{B}$)
they are less pronounced and the problem can be, in the first approximation,
treated as a long cylinder, since $\mathrm{P\ll D}$.

\begin{figure}[H]
\includegraphics[scale=0.4]{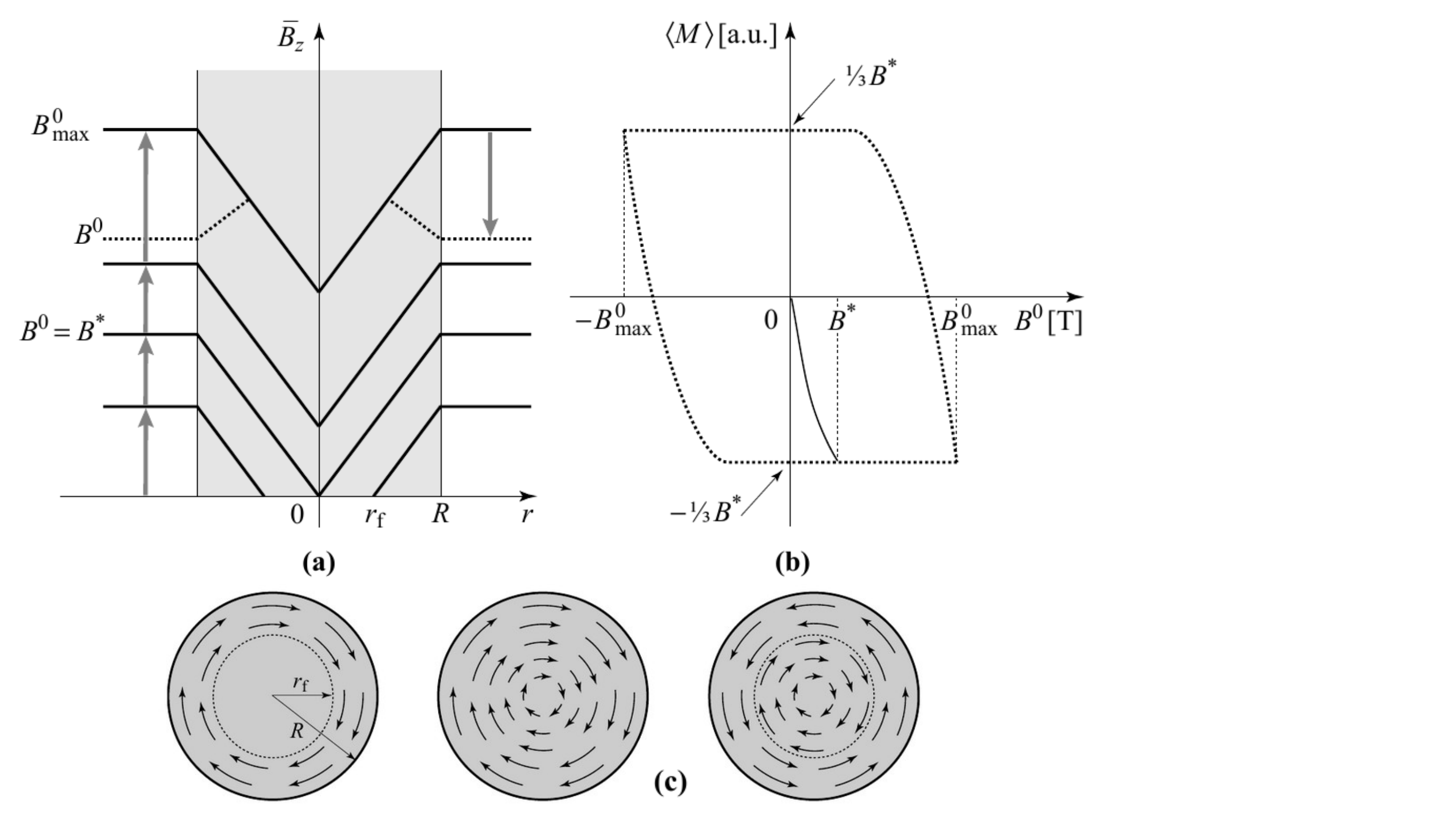}

\caption{The Bean critical state model for a long cylinder with radius $\mathrm{R\ll L}$:
$(a)$ distribution of local averaged field $\mathit{\mathrm{\overline{B}(r)}}$
in the external field $\mathrm{B^{0}}$ parallel to the surface of
the cylinder - $\mathit{up}$ $\mathrm{arrow}$ for increased $\mathrm{B^{0}}$
from $\mathrm{0}$ to $\mathrm{B_{max}^{0}}$ and $\mathit{down}$
$\mathit{arrow}$ for decreased $\mathrm{B^{0}}$ from $\mathrm{B_{max}^{0}}$
to $\mathrm{B^{0}}.$ For $\mathrm{B^{0}=B^{*}}$ the field penetrates
to the center of the sample, where $\mathit{\mathrm{\overline{B}(0)=0}}$;
$(b)$ the hysteresis loop for the magnetization $\mathrm{M_{ir}}=<\mathrm{\overline{M}}>$
(averaged over the cylinder) - the virgin (initial) line in $\mathit{bold}$;
(c) distribution of the current density in the superconducting cylinder
in an axial field: $\mathit{left}$ - in the increasing film from
$\mathrm{B^{0}=0}$ to $\mathrm{B^{0}<B^{*}}$ the screening current
flows between $\mathrm{r_{f}}$ and $\mathrm{R}$ where $\mathit{\mathrm{\overline{B}(r_{f})=0}}$;
$\mathit{middle}$ - for $\mathrm{B^{0}>B^{*}}$ it flows in the whole
sample; $\mathit{right}$ - in decreasing field from $\mathrm{B_{max}^{0}}$
to $\mathrm{B}^{0}$ the current is inverted in the external sheath.
From \cite{Mangin}.}
\end{figure}

\subsection{Theory for the penetration of the magnetic field in $\mathrm{H_{3}S}$
- perpendicular geometry}

In the perpendicular geometry $\mathrm{\mathbf{B}_{0}(=\mu_{0}\mathbf{H}_{0})\perp D}$
of the thin disk with $\mathrm{L\ll R}$ (i.e. $\mathrm{P\ll D}$
in the experiment of Ref. \cite{Troyan} - see $\mathit{Fig.}$ $\mathit{2}$)
there is a geometric barrier when the sample shape (due to geometric
spikes) is different from an ellipsoid. It turns out, that in that
case the magnetic field penetrates into the sample in an inhomogeneous
way \cite{Brandt}. The theory \cite{Brandt} predicts that in the
perpendicular geometry the flux is penetrated up the center in the
external field $\mathrm{B_{0\perp}=B_{p\perp}}$ with

\begin{equation}
B_{p\perp}(T)=B^{*}(T)\frac{P}{D}\mathrm{ln}\left[\frac{D}{P}+\sqrt{(1+\frac{D^{2}}{P^{2}}}\right],\label{eq:hperp}
\end{equation}
where $\mathrm{B^{*}(T)=\mu_{0}j_{c}(T)D/2}$. In order to explain
the temperature dependence of the field in the $^{119}$Sn sensor
\cite{Troyan} - see $\mathit{Fig.}$ $\mathit{4A}$, we define an
$\mathit{effective}$ $\mathit{field}$ on the $^{119}$Sn sensor
\begin{equation}
B_{Sn}^{\perp}(T)\equiv B_{0}-B_{p\perp}(T).
\end{equation}
(Here, $\mathrm{T}$ is the absolute temperature.) By definition,
for $\mathrm{\mathrm{\mathit{B_{\mathrm{0}}\leq B_{p\perp}(T)}}}$
one has $\mathrm{\mathrm{\mathit{\mathrm{B}_{Sn}^{\perp}=\mathrm{0}}}}$,
i.e. the field does not reach the center of the sample. When, $\mathrm{T\ll T_{c}}$
, $\mathrm{j_{c0}\approx(1.4-1.5)\times10^{7}}\textrm{Å}/\mathrm{cm^{2}}$
- see $\mathit{Fig.}$ $\mathit{6A}$, and $\mathrm{(P/D)\approx1/6}$
- see $\mathit{Fig.}$ $\mathit{2}$, one has $\mathrm{B_{p\perp}\approx1}$$\mathrm{T}$
which gives that
\begin{equation}
\mathrm{B_{0}(=0.68}T\mathrm{)<}B_{\mathrm{p\perp}}.
\end{equation}
This means, that the external field is not strong enough to push the
pinned vortices to the center of the sample. However, in order to
study the temperature dependence of $\mathrm{\mathit{B_{Sn}^{\perp}(T)}}$
it is necessary to know the temperature dependence of the current
density $\mathrm{\mathit{\mathrm{j}_{c}\mathrm{\mathrm{(}T})}}$.
In that respect, one should take into account two effects: ($\mathit{a}$)
The (mean field) temperature dependence of $\lambda(T)$ and $\xi(T)$
since $\mathit{\mathrm{j_{c}(T)}\sim\mathrm{\lambda^{-2}(T)\xi^{-1}(T)}}$.
In the temperature range $\mathrm{0<T\apprle T_{c}/2}$ we mimic $\mathrm{\mathrm{\mathit{\mathrm{\lambda^{-2}(T)\xi^{-1}(T)}\sim\mathrm{(1-T^{2}/T_{c}^{2})^{3}}}}}$;
($\mathit{b}$) The temperature fluctuations cause depinning effects
(see $\mathbf{II.C}$ ) which decrease $\mathit{\mathrm{j}_{c}\mathrm{(T)}}$
additionally - even bringing it almost to zero at some $\mathrm{T_{dp}^{r}}$
- see $\mathit{Fig.}$ $\mathit{1}$ and $\mathit{Fig.}$ $\mathit{6}$.
For the sake of simplicity, $\varphi\mathrm{(T)}$ is described approximately
by the linear function $\mathrm{\mathrm{\varphi(T)}\approx(T_{dp}^{r}-T)/(T_{dp}^{r}-T_{kink})}$
for $\mathit{T\geq T_{kink}}$. As the result, one obtains $\mathrm{\mathit{\mathrm{j_{c}(T)=j_{c0}\cdot\varphi(T)}\mathrm{(1-T^{2}/T_{c}^{2}}})^{3}}$.
The fit of the experimental curve in $\mathit{Fig.}$ $\mathit{6A}$
is obtained for $\mathit{\mathrm{T_{kink,\perp}\approx101}}$$\mathit{\mathrm{K}}$
and$\mathrm{\mathrm{\mathit{T_{dp,\perp}^{r}\approx\mathrm{122}}}}$$\mathit{K}$
which gives the theoretical curve for $\mathrm{\mathit{B_{Sn}^{\perp}(T)}}$
shown in $\mathit{Fig.}$ $\mathit{6A}$. 

It is important to point out, that the theoretical curve $\mathrm{\mathit{B_{Sn}^{\perp}(T)}}$
(in $\mathit{Fig.}$ $\mathit{6A}$) fits the experimental results
(in $\mathit{Fig.}$ $\mathit{4A}$) if the current density is high,
i. e. $\mathrm{\mathrm{\mathrm{\mathrm{j_{c0}}\approx\mathrm{1.4\times10^{7}}}\mathit{\mathrm{A/cm^{2}}}}}$.
This value for $\mathrm{j_{c0}}$ is compatible with the one obtained
from the $\mathit{TBR}$ measurements in $\mathrm{H_{3}S}$, where
$\mathrm{\mathit{\mathrm{j_{c0}}\gtrsim\mathrm{10^{7}}}\mathit{\mathrm{A}/\mathrm{cm^{2}}}}$.
Both findings confirm the assumption of the SCPC model, that the intrinsic
defects in $\mathrm{H_{3}S}$ are approximately elongated columns,
which pin vortices strongly, thus giving high critical current densities
in this material.

\begin{figure}
\includegraphics[scale=0.35]{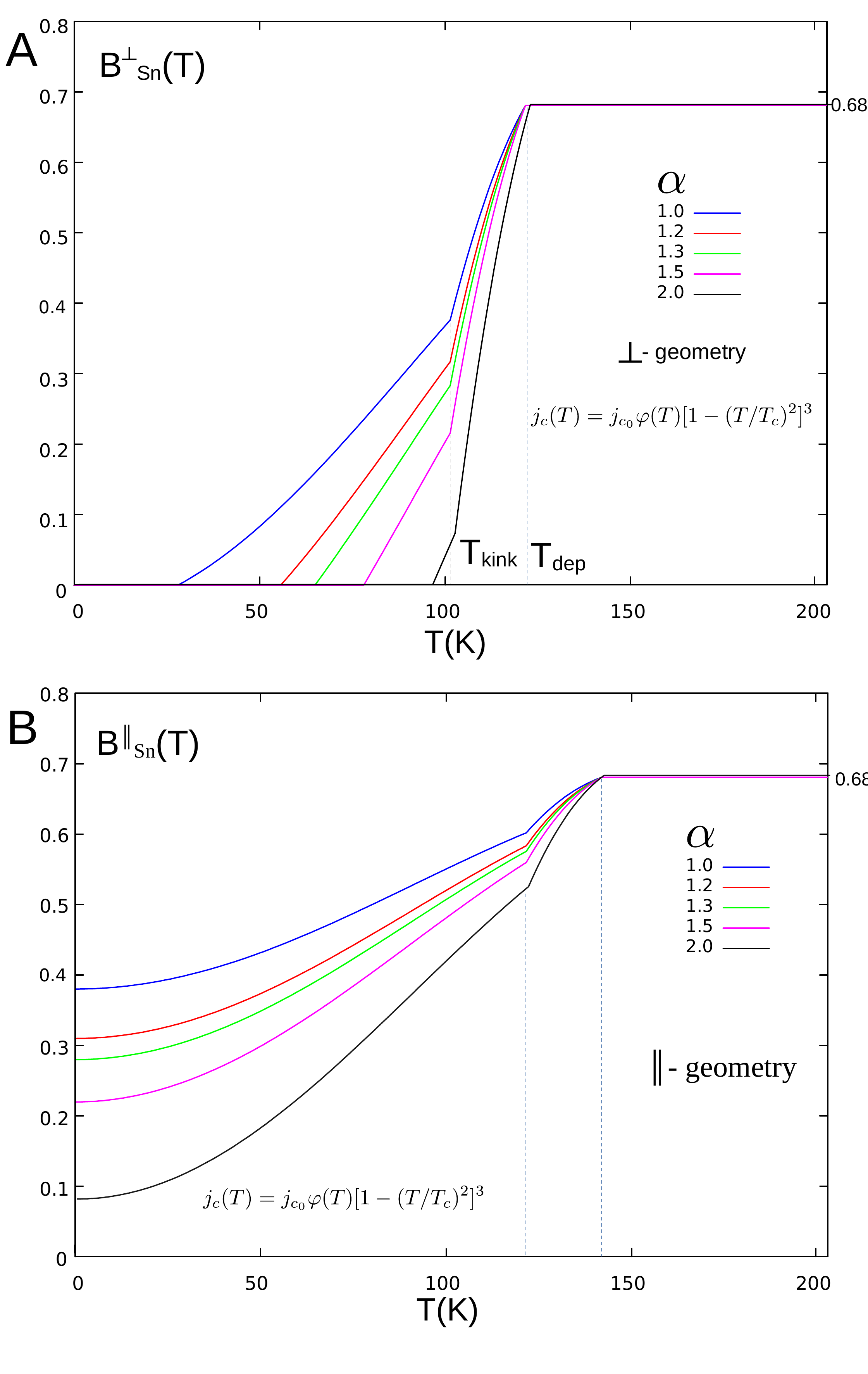}

\caption{The temperature dependence of the effective magnetic field $\mathit{B_{Sn}(T)}$
in the $^{119}$Sn sensor in the combined Bean model and the SCPC-model.
For $\mathrm{T_{kink}}$ and $\mathrm{T_{dep}(\equiv T_{dp}^{r})}$
see text. $\mathbf{A}$ - $\mathit{B_{Sn}^{\perp}(T)}$ in the perpendicular
geometry; $\mathbf{B}$ - $\mathit{B_{Sn}^{\parallel}(T)}$ in the
parallel geometry. The various curves are for different current densities
$\mathrm{\mathrm{\mathit{\mathrm{j_{c0}}=\alpha\times\mathrm{10^{7}}}}\mathit{A/cm^{2}};\mathrm{\alpha}=\mathrm{\mathit{\mathrm{1},\mathrm{1.2}},\mathit{\mathrm{1.3},\mathrm{1.5},\mathrm{2.0}}}}$}

\end{figure}

\subsection{Theory for the penetration of the magnetic field in $\mathrm{H_{3}S}$
- parallel geometry}

Similarly, when the field is parallel to the thin disk, i.e. $\mathrm{\mathit{B^{\parallel}(T)\parallel D}}$
- see $\mathit{Fig.}$ $\mathit{1}$, the effective field on the $^{119}$Sn
sensor is given by
\begin{equation}
B_{Sn}^{\parallel}(T)\equiv B_{0}-B_{p\parallel}(T).
\end{equation}
In this geometry and with $\mathrm{P\ll D}$ - see $\mathit{Fig.}$
$\mathit{2}$, the system can be approximated by a long cylinder with
$\mathrm{R\approx P/2}$ and $\mathrm{L\approx D}$. In that case
the formula Eq.(\ref{eq:field}) holds approximately, which gives
$\mathrm{\mathit{\mathrm{B_{p\parallel}(T)\approx}\mathrm{\mu_{0}j_{c}(T)P/2}}}$.
For assumed value $\mathrm{\mathrm{\mathrm{\mathrm{j_{c0}}\approx\mathrm{1.4\times10^{7}}}}}$
$\mathrm{A/cm^{2}}$one obtains $\mathrm{B_{p\parallel}(T\ll T_{c})\approx0.45}$
$\mathrm{T}$, which is smaller than the applied field $\mathrm{B_{0}=0,68}$
$\mathrm{T}$, i.e. 
\begin{equation}
\mathrm{B_{p\parallel}(T\ll T_{c})<B_{0}(=0,68}\mathrm{T})
\end{equation}
This means, that $\mathrm{B_{Sn}^{\parallel}(T\ll T_{c})>0}$ and
the external field $\mathrm{B_{0}}$ is large enough to push vortices
into the center of the $^{119}$Sn film - see $\mathit{Fig.}$ $\mathit{6B}$,
which is in agreement with the experimental result in $\mathit{Fig.}$
$\mathit{4B}$. The fit of $\varphi(T)$ (note, that $\mathrm{j_{c}(T)\sim\varphi(T)}$)
is analogous to the perpendicular case but with slightly different
parameters $\mathrm{T_{kink,\parallel}\approx122}$$\mathrm{K}$ and
$\mathrm{T_{dp,\parallel}^{r}\approx142}$$\mathrm{K}$ . Note, that
there is a qualitative difference between $\mathrm{B_{Sn}^{\perp}(T)}$
and $\mathrm{B_{Sn}^{\parallel}(T),}$ since $\mathrm{B_{Sn}^{\perp}(T)}$
is zero up to some finite temperature - see $\mathit{Fig.}$ $\mathit{4}$
and $\mathit{Fig.}$ $\mathit{6}$, and $\mathrm{B_{Sn}^{\parallel}}(T)$
is finite even at $\mathrm{T=0}$$\mathrm{K}$. This is due to different
sample dimensions, i.e. that $\mathrm{P\ll D}$ - see $\mathit{Fig.}$
$\mathit{2}$. These results are in a qualitative and semi-quantitative
agreement with the experimental results in \cite{Troyan}. 

It is necessary to point out following items: ($\mathit{a}$) It is
seen in $\mathit{Fig.}$ $\mathit{5}$, that in the Bean critical
state model the irreversible magnetization for a long cylinder in
fields $\mathrm{B_{0}>B^{*}}$ is constant (field independent). Since
the measurements of the magnetic moment of the $\mathrm{H_{3}S}$
sample in \cite{Troyan} are given in the CGS unit $\mathrm{emu}$
we use it also here. After averaging of $\mathrm{\overline{M}=(\overline{B}-H)/4\pi}$
over the sample one has $\mathrm{\overline{M}_{ir}(emu/cm^{3})\approx}(R/30)\mathrm{j}_{c}(\mathrm{A/cm^{2}})$
and the total magnetic moment $\mathrm{\mu}_{\mathrm{ir}}=$$\bar{\mathrm{M_{ir}}}\cdot\mathrm{\mathrm{V_{s}}}$.
Note, that $\mathrm{\bar{M}_{ir}=(\bar{M}_{+}-\bar{M}_{-})/2=\bar{M}_{-}}$
since in the Bean model $\mathrm{\bar{M}_{+}=-\bar{M}}$ holds for
$\mathrm{B_{0}>B^{*}}$. For the volume of the disk $\mathrm{V_{s}\approx0.8\times P\cdot D^{2}}$
one obtains the trapped magnetic moment in the sample to be of the
order $\mu\approx(0.3-2)\times10^{-5}$ $\mathrm{emu}$ for $\mathrm{B_{0}>B_{p\perp}}$
and for $\mathrm{j_{c0}\approx(1.4-10)\times10^{7}}$$\mathrm{A/cm^{2}}$.
The calculated $\mathit{trapped}$ magnetic moment in the $\mathrm{H_{3}S}$
sample of Ref.\cite{Troyan} is far beyond the SQUID sensitivity threshold
- which is $\sim10^{-8}$ $\mathrm{emu}$ \cite{Troyan}. This means
that the measurements of the trapped magnetic flux of vortices (but
without the extrinsic moments) are realizable. ($\mathit{b}$) In
order to explain the field penetration in the $\mathrm{H_{3}S}$ sample
of Ref. \cite{Troyan} it comes out that the perpendicular ($\mathbf{H\perp}$$\mathrm{D}$)
critical current density $\mathrm{j_{c0}^{\perp}}$ must be approximately
equal to the parallel ($\mathbf{H\parallel\mathrm{D}}$) one $\mathrm{j_{c0}^{\parallel}}$,
i. e. $\mathrm{j_{c0}^{\perp}}\approx\mathrm{j_{c0}^{\parallel}}$.
This means, that in the $\mathrm{H_{3}S}$ sample of Ref. \cite{Troyan}
the columnar defects are oriented along both directions, the parallel
and perpendicular one, in a similar way - schematically shown in $\mathit{Fig.}$
$\mathit{2}$. Having in mind the cubic-like structure of $\mathrm{H_{3}S}$
this kind of pinning isotropy is an acceptable assumption; ($\mathbf{c}$)
In order to explain the $\mathrm{\mathit{TBR}}$ and $\mathit{PMF}$
measurements in thin $\mathrm{H_{3}S}$ samples in the SCPC model,
it comes out that the bulk $\mathrm{H_{3}S}$ sample is a high-$\kappa$
type II superconductor with the parameters: $\xi_{0}\approx(15-20)$
$\textrm{Å}$, $\lambda_{0}\approx(1-2)\times10^{3}$$\textrm{Å}$
; $\kappa\approx(50-100)$, $\mathrm{\mu_{0}H_{c1}}\approx(18-60)$
$\mathrm{m\mathrm{T}}$, $\mathrm{\mathrm{\mu_{0}}H_{c0}}$$\mathrm{\approx(0.6-1.1)}$
$\mathrm{\mathrm{T},}$ $\mathrm{\mathrm{\mu_{0}}H_{c2}}\approx(80-140)$
$\mathrm{T}$. These values are very different from those obtained
in \cite{Eremets-Meissner-2022}, \cite{Minkov-Meissner-2021}, where
$\xi_{0}\sim20$ $\textrm{Å}$, $\lambda_{0}\sim(1.3-2)\times10^{2}$$\textrm{Å}$
; $\kappa\sim7-10$, $\mathrm{\mathrm{\mu_{0}}H_{c0}}$$\mathrm{\sim6}$
$\mathrm{\mathrm{T},}$ $\mathrm{\mathrm{\mu_{0}}H_{c1}\sim1}$ $\mathrm{T}$,
$\mathrm{\mu_{0}H_{c2}}\approx(80-140$) $\mathrm{T}$. The values
of the parameters predicted in the SCPC model are compatible with
those obtained in the magnetic measurements and also with the microscopic
theory - see the discussion below.

To this point, Hirsch and Marsiglio have recently realized \cite{Hirsch-optics-arXiv},
that their previous interpretation \cite{Hirsch-1-1}-\cite{Hirsch-2-1}
of the Troyan's measurements in $\mathrm{H_{3}S}$ \cite{Troyan}
in terms of pinning-free superconductors with unphysically high $\mathrm{j_{c0}}$$\sim10^{11}\mathrm{A}/\mathrm{cm^{2}}$
may be inadequate. Namely, due to the pronounced magnetization hysteresis
in \cite{Drozdov-2015} they speculated the presence of pinning forces
in $\mathrm{H_{3}S}$ - with the critical current density $\mathrm{j_{c0}}\sim10^{7}\mathrm{A}/\mathrm{cm^{2}}$.
However, they did not realize that the $\mathit{TBR}$ and $\mathit{PMF}$
effects in $\mathrm{H_{3}S}$ are due to the strong pinning by the
elongated (columnar) defects - as the SCPC model predicts \cite{Kulic-RTSC-v1}.

To conclude this Section - the magnetic measurements of the penetration
of the magnetic field $B_{0}=0.68$ $\mathrm{T}$ in the $\mathrm{H_{3}S}$
sample \cite{Troyan}can be naturally explained in the framework of
the SCPC model and the Bean critical state model. This approach also
explains naturally the high critical current density in the $\mathrm{H_{3}S}$
samples, which is of the order $\mathrm{j_{c0}\approx(1.4-1.5)\times10^{7}}$
$\mathrm{A/cm^{2}}$. Thereby, the finite-size effects in the thin
$\mathrm{H_{3}S}$ disk of Ref. \cite{Troyan} are taken into account.
This analyzes tells us, that there is no need to call into question
the existence of superconductivity in $\mathrm{H_{3}S}$ (and in other
HP-hydrides), as it is claimed in \cite{Hirsch-1-1}-\cite{Hirsch-Marsiglio-arXiv V5-2022-1}.

\subsection{Meissner effect in $\mathrm{H_{3}S}$}

The Meissner effect is an important hallmark of the superconducting
state. It is realized in the so called $\mathrm{FC}$ ($\mathit{field}$
$\mathit{cooled}$) $\mathit{experiment}$, when the sample (ideally
without pinning defects) is in the normal state ($\mathrm{T>T_{c}}$)
and placed in a magnetic field $\mathrm{H_{0}<H_{c1}}$(for type-II
superconductors). The latter penetrates into the normal metallic sample
fully, i. e. one has $\mathit{B\approx\mu_{0}\mathrm{H_{0}}}$. However,
if the sample is then cooled down into the superconducting state ($\mathrm{T<T_{c}}$)
the magnetic field will be $\mathit{expelled}$ from the bulk sample,
i. e. one has $\mathrm{\overline{B}=0}$ in an ideal non-magnetic
superconductor. The Meissner effect should not be confused with the
$\mathrm{ZFC}$ ($\mathrm{\mathit{\mathrm{zero}-\mathrm{f}\mathrm{ield}\mathrm{\textrm{ }cooled}}}$)
$\mathit{experiment}$, where a nonmagnetic bulk metallic sample is
first cooled into the superconducting state at $\mathrm{T<T_{c}}$
in the zero field ($\mathrm{H=0}$) and thereafter magnetic field
$\mathrm{H<H_{c1}}$ is turned on. As a result, the magnetic field
(induction) is $\mathit{excluded}$ from the bulk sample, i.e. $\mathrm{B=\mu H=0}$
with $\mu=\mu_{0}(1+\chi)=0$. The $\mathrm{ZFC}$ phenomenon is also
called $\mathit{diamagnetic}$ $\mathit{shielding}$. However, the
$\mathrm{ZFC}$ effect would also be realized in a $\mathit{perfect}$
$\mathit{metal}$ (with $\varrho=0$) - if it existed in nature, where
it is due to the classical Lenz law of the electrodynamics. From this
analyzes comes out, that the magnetic susceptibility of an ideal bulk
superconductor is diamagnetic $\chi=-1$ in the SI system ($\mathrm{4\pi\chi=-1}$
in the $\mathrm{CGS}$) in both types of experiments. Note, that when
the FC experiment is done in a perfect metal, the magnetic field is
not expelled from the sample at $\mathrm{T<T_{c}}$, i.e. the magnetic
flux stays frozen in the sample with the same value as in the normal
state, i.e. $\mathrm{\mathit{\mathrm{B}\approx\mu_{0}\mathrm{H_{0}}}}$.
So, for a definite proof of the Meissner effect in superconductors
one should perform the FC experiment. 
\begin{figure}
\includegraphics[scale=0.41]{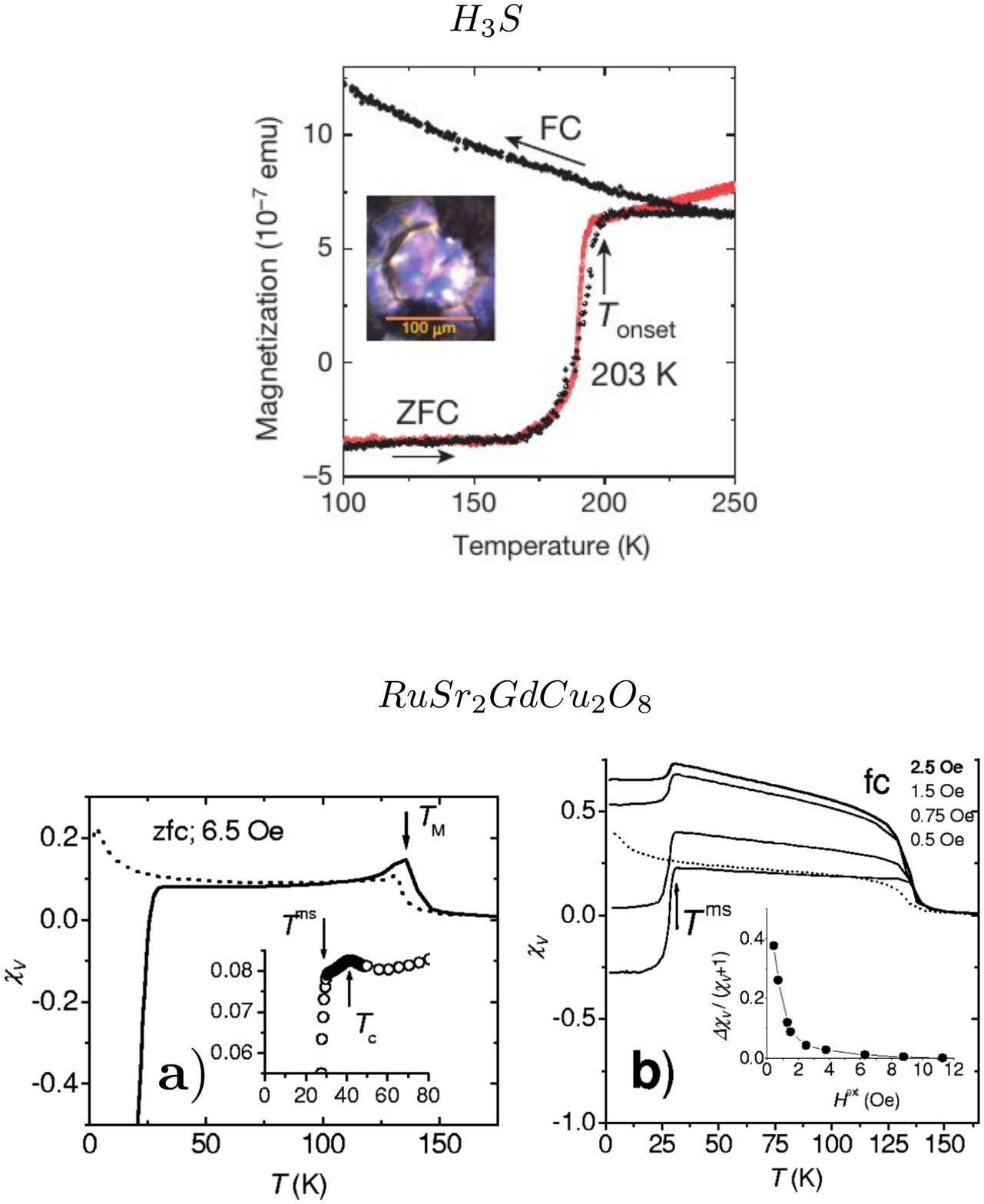}

\caption{$\mathit{Up}$ - The $\mathrm{FC}$ and $\mathrm{ZFC}$ magnetization
($\mathrm{M}$) in $\mathrm{H_{3}S}$ - from \cite{Eremets-Meissner-2022},\cite{Drozdov-2015}.
$\mathit{Down}$ - the volume susceptibility $\chi_{\mathrm{V}}$
of the weak ferromagnetic superconductor $\mathrm{RuSr_{2}GdCu_{2}O_{8}}$;
$\mathbf{a)}$ - ZFC (zfc) measurements of $\chi_{\mathrm{V}}$at
$\mathrm{H^{ex}=6.5}$ $\mathrm{Oe}$; doted line is $\chi_{V}$ in
the non-superconducting state. $\mathrm{T_{M}\approx137}$ $\mathrm{K}$
is the magnetic critical temperature; $\mathrm{T^{ms}\approx30}$
$\mathrm{K}$ is the transition temperature to the spontaneous vortex
state ($\mathrm{SVS}$) which exists at $\mathrm{T^{ms}<T<T_{c}}$.
Inset: enlarged scale of $\chi_{\mathrm{V}}$ around the superconducting
transition $\mathbf{\mathrm{T_{c}}\approx\mathrm{45}}$$\mathrm{K}$
; $\mathbf{b)}$ FC measurements of $\chi_{\mathrm{V}}$ for external
fields $H^{ex}\approx(0.5-2.5)$ $\mathrm{Oe}$. Inset: The volume
fraction of the Meissner state $\mathrm{f}\approx40\%$ at $\mathrm{H^{ex}=0,5}$$\mathrm{Oe}$
- from \cite{Tallon-Ru}.}

\end{figure}

In that respect, several inconsistent experimental and theoretical
results related to the magnetization measurements in $\mathrm{H_{3}S}$
(and $\mathrm{LaH_{10}}$) were published (in the period 2015-2022)
in \cite{Eremets-Meissner-2022},\cite{Drozdov-2015},\cite{Minkov-Meissner-2021}.
These results are strongly criticized in \cite{Hirsch-1-1}-\cite{Hirsch-Marsiglio-arXiv V5-2022-1}.
We shall not elaborate this criticism in details, but enumerate few,
in our opinion, main points by Hirsch and Marsiglio \cite{Hirsch-1-1}-\cite{Hirsch-Marsiglio-arXiv V5-2022-1}.
These are: $\mathit{\mathrm{(\mathbf{\mathrm{\mathit{1}}})}}$ The
Meissner effect is not observed experimentally in $\mathrm{H_{3}S}$
(and other HP-hydrides) \cite{Eremets-Meissner-2022},\cite{Drozdov-2015},\cite{Minkov-Meissner-2021}).
Namely, in the FC measurements only some parasitic paramagnetic susceptibility
is observed, i. e. $\chi_{FC}(\mathrm{T<T_{c}})>0$, and there are
no signs of the flux expulsion. However, in the ZFC experiment there
is a clear diamagnetic shielding (at $\mathrm{T<T_{c}}$) but existing
on the background of the parasitic paramagnetic signal. Most probably,
the origin of this paramagnetic signal is not intrinsic, since there
is no reliable reason for strong paramagnetic effects in $\mathrm{H_{3}S}$.
To this point, there is also a pronounced paramagnetic contribution
in the FC measurements in the HTSC superconductor $\mathrm{RuSr_{2}GdCu_{2}O_{8}}$
- with $\mathrm{T_{c}}$$\approx45$ $\mathrm{K}$, where a weak ferromagnetism
appears below $\mathrm{T_{M}\approx137}$ $\mathrm{K}$ \cite{Tallon-Ru}.
At $\mathrm{T\geq T^{ms}\approx30}$ $\mathrm{K}$ a spontaneous vortex
state ($\mathrm{SVS}$) appears, where $\mathrm{H_{c1}(T)<M}$ and
$\mathrm{M}$ is the spontaneous magnetization. In $\mathit{Fig.}$
$\mathit{7}$ it is seen, that in the FC measurements $\chi_{v}$
is paramagnetic, even for $\mathrm{T<T_{c}}$, i. e. $\chi_{\mathrm{V}}(T)>0$
. At $\mathrm{T^{ms}\approx30}$ $\mathrm{K}$ the susceptibility
$\chi_{V}$ decreases suddenly and the sample is in the bulk Meissner
state at $\mathrm{T<T^{ms}}$ with the volume fraction $\mathrm{f=\mid(\chi_{V}(0)-\chi_{V}(T^{ms})/(1+\chi_{V}(T^{ms}))\mid\approx40}\%$
at $\mathrm{H^{ex}}=0,5$ Oe - see Inset in $\mathit{Fig.}$ $\mathit{7b}$.
This tells us that pinning centers are present in the $\mathrm{RuSr_{2}GdCu_{2}O_{8}}$
sample. 

Having in mind this analyzes, one expects that in the FC measurements
in $\mathrm{H_{3}S}$ a sudden decrease of $\chi_{V}$ should be realized
below $\mathrm{T_{c}}$. However, so far, no such results have been
published with clearly realized Meissner effect in HP-hybrides. The
paramagnetic signal in the $\mathrm{H_{3}S}$ sample is most probable
due to some extrinsic properties of the diamond anvil cell (or of
the rest of the sulfur atoms, since $\mathrm{H_{2}S}$ is decomposed
into $\mathrm{H_{3}S}$ and $\mathrm{S}$) and this fact deserves
further studies. $\mathrm{(\mathit{2})}$ Additional critical remarks,
given in \cite{Hirsch-1-1}-\cite{Hirsch-Marsiglio-arXiv V5-2022-1},
are related to some controversial findings in \cite{Minkov-Meissner-2021},
that in the $\mathrm{H_{3}S}$ (and $\mathrm{LaH_{10}}$) sample the
critical fields $\mathrm{H_{c1}(0)}$ and $\mathrm{H_{c}(0)}$ are
too large, while the penetration depth $\lambda_{0}$ is too small.
If this is true, the large value of the critical field $\mathrm{\mathrm{\mu_{0}}H_{c1}(0)}$
in \cite{Minkov-Meissner-2021} would give very high critical current
density in $\mathrm{H_{3}S}$ (and $\mathrm{LaH_{10}}$), i. e. $\mathrm{j}_{c0}\sim10^{10}$
$\mathrm{A/cm^{2}}$. This is much higher value than the depairing
current density $\mathrm{j_{dep}\simeq5\times10^{8}}$$\mathrm{A/cm^{2}}$,
what is in fact impossible. The large value for $\mathrm{H_{c}(0)}$
(in Ref. \cite{Minkov-Meissner-2021}) is incompatible with the microscopic
theory of superconductivity, which relates the condensation energy
to $\mathrm{H_{c}(0)}$ by $\mathrm{(N(E_{F})\Delta^{2})=H_{c}^{2}(0)/8\pi}$.
Here, $\mathrm{N(E_{F})}$ is the density of states (per unit volume)
at the Fermi surface. For $\mathrm{2\Delta\approx3,5}$$\mathrm{T_{c}}$
and $\mathrm{\mathrm{\mu_{0}}H_{c}(0)\sim10}$ $\mathrm{T}$ (in \cite{Minkov-Meissner-2021})
one obtains that $\mathrm{N(E_{F})}\approx$$\mathrm{30\times N^{DFT}(E)}$,
while the density functional theory gives $\mathrm{N^{DFT}(E_{F})\approx0,2}$
$\mathrm{states/spin\times eV\times}\textrm{(Å})^{3}$ \cite{Hirsch-1-1}-\cite{Hirsch-Marsiglio-arXiv V5-2022-1}.
Therefore, this highly overestimated value for the density of states,
which are extracted from the magnetic measurements in $\mathrm{H_{3}S}$
\cite{Eremets-Meissner-2022},\cite{Drozdov-2015},\cite{Minkov-Meissner-2021},
is an highly unacceptable value. 

Let us discuss the origin of these too large values for the bulk critical
fields $\mathrm{\mathrm{\mu_{0}}H_{c1}(\sim1}$ $\mathrm{T})$ and
$\mathrm{\mathrm{\mu_{0}}H_{c0}(\sim10}$ $\mathrm{T)}$ obtained
in \cite{Minkov-Meissner-2021}. The latter result is based on an
inadequate experimental definition of $\mathrm{H_{c1}}$. Namely,
it is determined from the onset field $\mathrm{\mathrm{\mu_{0}}H_{p}(0)(\sim0.1}$$\mathrm{T)}$
of the deviation of $\mathrm{M(H)}$ from the linear dependence by
assuming that $\mathrm{H_{c1}=H_{p}(0)/(1-N)}$. In the measurements
in Ref. \cite{Minkov-Meissner-2021} the demagnetization factor is
$\mathrm{N\approx0.96}$, what gives too large value for $\mathrm{\mathrm{\mu_{0}}H_{c1}(T=0)\approx2}$
$\mathrm{T}$. That this procedure is not well defined, i. e. $\mathrm{H_{p}(0)}$
is not related to $\mathrm{H_{c1}}$, can be seen in $\mathit{Fig.}$
$\mathit{5b}$ where $\mathrm{M(H)}$ curve in the Bean critical state
model, where the saturation field $\mathrm{H^{*}(P,D)\gg H_{c1}(0)}$.
In the perpendicular geometry of the experiment \cite{Troyan} one
has $\mathrm{P=5}$ $\mathrm{\mu m}$ and $\mathrm{D=30}$ $\mathrm{\mu m}$
, $\mathrm{N\sim0.7}$ and according to $\mathit{Eq.}$ $\mathit{(9)}$
one has $\mathrm{\mathrm{\mu_{0}}H_{p,\perp}(0)\approx1}$ $\mathrm{T}$
what is much larger than the real $\mathrm{H_{c1}(T=0)}$. If we would
apply the same procedure for obtaining $\mathrm{H_{c1}}$ in the $\mathrm{H_{3}S}$
sample, as it was done in Ref. \cite{Minkov-Meissner-2021}, we would
obtain also an unrealistic value $\mathrm{\mathrm{\mu_{0}}H_{c1}(T=0)\approx3}$
$\mathrm{T}$ , instead of the realistic one $\mathrm{\mathrm{\mu_{0}}H_{c1}(0)}\approx(18-60)$
$\mathrm{m\mathrm{T}}$.

\section{Summary and discussion }

Recently, the authors o Refs. \cite{Hirsch-1-1}-\cite{Hirsch-Marsiglio-arXiv V5-2022-1}
raised important questions on the reliability of magnetic measurements
in high-pressure hydrides (HP-hydrides). Their skepticism goes so
far, that they tend to conclude that superconductivity does not actually
exist in HP-hydrides \cite{Hirsch-1-1}-\cite{Hirsch-Marsiglio-arXiv V5-2022-1}.
This attitude is mostly related to the FC magnetic measurements, which
are until now unable to prove unambiguously the existence of the Meissner
effect in small samples of $\mathrm{H_{3}S}$. On the other side,
some ZFC measurements, in small samples of $\mathrm{H_{3}S}$, are
experimentally more reliable, because these are not related to the
flux trapping effects. There are also difficulties to explain some
experimental results of magnetic measurements in $\mathrm{H_{3}S}$
- done in \cite{Eremets-Meissner-2022},\cite{Drozdov-2015},\cite{Minkov-Meissner-2021},
if they are treated by the standard theory of superconductivity with
weak pinning of vortices. The latter approach is mainly accepted in
\cite{Hirsch-1-1}-\cite{Hirsch-Marsiglio-arXiv V5-2022-1}, thus
reducing the possibility for explaining experiments in $\mathrm{H_{3}S}$,
such as: ($\mathit{i}$) $\mathrm{TBR}$ - the temperature broadening
of resistance in magnetic field, and ($\mathit{ii}$) $\mathit{PMF}$
- the penetration of the magnetic field into the center of the sample. 

In order to explain these two kinds of experiments - which can not
be explained by the weak pinning theory, Ref. \cite{Kulic-RTSC-v1}
introduces the SCPC model - which holds for superconductors with strong
and elongated pinning defects. Moreover, even the quantitative explanation
of these two phenomena (in $\mathrm{H_{3}S}$ ) is possible by assuming
that the pinning centers are in the form of long columnar defects,
which are ``isotropically'' distributed over the sample, i. e. with
the same values of the critical current densities flowing perpendicular
and parallel to the sample surface. In the framework of the SCPC model
it is possible to explain these two kind of experiments. In the following
we summarize the obtained results: $\mathbf{1}$. The large reduction
of $TBR$, $\mathrm{\mathrm{\delta t_{c}^{scpc}(h)}}$, is due to
the long columnar defects $\mathrm{L}$$\approx\mathrm{L_{v(ortex)}}$$\gg\xi_{0}$
with the radius $\mathrm{r}\sim\xi_{0}$. In such a case, both the
core and the electromagnetic pinning are operative. These cause high
densities of the critical current (at $\mathrm{T\ll T}_{c}$) - of
the order $\mathrm{j_{c0}=(10^{7}-10^{8})}$ $\mathrm{A/cm^{2}}$.
The SCPC model predicts, that in $\mathrm{H_{3}S}$ the temperature
width of $\mathit{TBR}$ is governed by the small parameter $\mathrm{C=}$$\xi_{0}/\mathrm{L_{col}}$,
i.e. $\mathrm{\mathrm{\delta t_{c}^{scpc}(h)}\sim C^{1/2}h^{1/2}}$
with $\mathrm{C\sim10^{-3}}$ and $\mathrm{h=H/H_{c2}}$. This gives
$\mathrm{\mathrm{\delta t_{c}^{scpc}(h)}\apprle0.01}$ for $\mathrm{h\lesssim0.01}$
and $\mathrm{L}\sim1$ $\mathrm{\mu m}$, which is in satisfactory
agreement with experimental results in $\mathrm{H_{3}S}$ \cite{Eremets-Meissner-2022},
as shown in $\mathit{Fig.}$ $\mathit{3}$. It is also seen that the
SCPC model fits the experimental results much better than the model
with weak pinning by small defects. The SCPC model also predicts,
that the magnetic irreversible field is governed by $\mathrm{C^{-1}}$,
i.e. $\mathit{\mathrm{H_{irr}\sim(1-t)^{2}L/\xi_{0}}}$. The irreversibility
line is not only significantly increased compared to HTSC-cuprates,
but the temperature dependence, as a measure of the strength of pinning
forces, is given by $\mathit{\mathrm{(1-t)^{2}}}$ instead of $\mathit{\mathrm{(1-t)^{3/2}}}$
- characteristic for materials with point-like defects. Measurements
of the irreversible line $\mathrm{H_{irr}}$ in $\mathrm{H_{3}S}$
are desirable.

These columnar defects cause high density of the critical current
and also a large magnetization hysteresis $\Delta M$ in $\mathrm{H_{3}S}$.
This property opens a possibility for making powerful high-field superconducting
magnets. For instance, by making (if possible) a long superconducting
cylinder of $\mathrm{H_{3}S}$ (with $\mathrm{L\gg R}$) the magnetic
hysteresis in that case is given by $\Delta M\sim\mathrm{j_{c0}\times R}$,
where $\mathrm{R}$ is the radius of the superconducting cylinder.
For instance, in $\mathrm{H_{3}S}$ for $\mathrm{j_{c0}}>10^{7}$$\mathrm{A/cm^{2}}$
and for $\mathrm{R\sim0.6}$ $\mathrm{cm}$ one has $\mathrm{\mathrm{\mu_{0}}M}\sim100$
$\mathrm{T}$ at the temperature $\mathrm{T<50}$ $\mathrm{K}$. 

$\mathbf{2.}$ $\mathrm{\mathit{PMF}}$ in $\mathrm{H_{3}S}$ can
be naturally explained by the SCPC model, where the strong columnar
pinning of vortices dominates. The experiment measures $\mathit{\mathrm{\mathit{PMF}}}$
in the applied field $\mathrm{\mathrm{B_{0}}=\mathrm{\mu}_{0}H_{0}=0.68}$
$\mathrm{T}$ \cite{Troyan}. If the field is penetrated in the center
of the sample, then the quantum beats should appear in the $^{119}$Sn
sensor. It turns out that in the perpendicular geometry, when the
field is perpendicular to the sample surface ($\mathrm{B_{0}\perp D}$
- see $\mathit{Fig.}$ $\mathit{2}$), the magnetic field does not
penetrate to the center, while in the parallel case ($\mathrm{B_{0}\parallel D}$
- see $\mathit{Fig.}$ $\mathit{2}$) it penetrates partially even
for $\mathrm{T\ll T_{c}}$ \cite{Troyan}. These experimental results
are naturally explained in the SCPC model, where the ``isotropically''
distributed strong columnar pinning defects make a large current density
$\mathrm{j_{c0}^{\perp}\approx j_{c0}^{\parallel}\approx(1.3-1.5)}$$\times10^{7}$
$\mathrm{A/cm^{2}}$. It is also predicted, that the depinning temperature
$\mathrm{T_{dp}^{r}}$ - where the critical current density is strongly
weakened ($\mathrm{j_{c}(T)\ll j_{c0}}$) and the field $\textrm{\ensuremath{\mathrm{B_{0}}}}$
is fully penetrated into the $\mathrm{H_{3}S}$ sample, is of the
order $\mathrm{T_{dp}^{r}\sim(100-120)}$$\mathrm{K}$.

Moreover, in order to explain the $\mathit{TBR}$ and $\mathit{PMF}$
experiments by the SCPC model it comes out that $\mathrm{H_{3}S}$
is a high-$\kappa$ superconductor with bulk physical parameters:
$\xi_{0}\approx(15-20)$ $\textrm{Å}$, $\lambda_{0}\approx(1-2)\times10^{3}$$\textrm{Å}$
; $\kappa\approx(50-100)$, $\mathrm{\mathrm{\mu_{0}}H_{c1}(0)\approx(18-60)}$
$\mathrm{m\mathrm{T}}$, $\mathrm{\mathrm{\mu_{0}}H_{c0}\approx}$$\mathrm{(0.6-1.1)}$
$\mathrm{\mathrm{T},}$ $\mathrm{\mathrm{\mathrm{\mu_{0}}H}}_{c2}\approx(80-140)$
$\mathrm{T}$. These values of parameters are also compatible with
the microscopic theory of superconductivity. Finally, it is natural
to raise the question - what is the origin of these columnar pinning
defects in $\mathrm{H_{3}S}$? Serious candidates are single edge
dislocations or their bundles, what is a matter of further research.
To conclude - the SCPC model, which assumes the existence of strongly
acting columnar pinning centers, is able to explain important $TBR$
and $\mathit{PMF}$ measurements in a high-$\kappa$ $\mathrm{H_{3}S}$
superconductor. Therefore, there is no need for calling into question
the existence of superconductivity in $\mathrm{H_{3}S}$ \cite{Hirsch-1-1}-\cite{Hirsch-Marsiglio-arXiv V5-2022-1}.
\begin{acknowledgments}
The author would like to thank Dirk Rischke and Radoš Gaji\'{c} for
permanent support and to Igor Kuli\'{c} for discussions and support.
\end{acknowledgments}

\end{document}